\documentclass[12pt]{article}
\usepackage{graphicx}
\usepackage{dcolumn}
\usepackage{bm}
\usepackage{amsmath, amssymb, amsfonts, amsthm}
\usepackage{cite, wrapfig}
\usepackage{epsfig}
\usepackage[usenames,dvipsnames,table]{xcolor}
\usepackage{textcomp}
\usepackage{tikz-cd} 
\usepackage{ dsfont }
\usepackage[utf8]{inputenc}
\usepackage{amsmath}
\usepackage{tikz}
\usepackage{youngtab}
\usepackage{hyperref}

\usepackage{colortbl}
\definecolor{summersky}{cmyk}{0.71,0.33,0,0.5}
\definecolor{flamingo}{cmyk}{0,0.51,0.71,0.5}
\definecolor{rp}{cmyk}{0.2, 1, 0.6, 0}
\definecolor{pacificblue}{cmyk}{0.95,0.3,0, 0.5}
\definecolor{gray60}{cmyk}{0.4,0.4,0,0.8}

\hypersetup{
    pdftoolbar=true,        
    pdfmenubar=true,        
    pdffitwindow=true,     
    pdfstartview={FitH},    
    pdfnewwindow=true,      
    colorlinks=true,       
    linkcolor=pacificblue,          
    citecolor=flamingo,        
    filecolor=magenta,      
    urlcolor=pacificblue           
}

\def\be{\begin{equation}}
\def\ee{\end{equation}}
\def\ba{\begin{eqnarray}}
\def\ea{\end{eqnarray}}

\def\bdm{\begin{displaymath}}
\def\edm{\end{displaymath}}

\def\bq{\begin{quote}}
	\def\eq{\end{quote}}

 at 10truept

\newcommand{\bea}{\begin{eqnarray}}
\newcommand{\eea}{\end{eqnarray}}

\newcommand{\bi}{\begin{itemize}}
	\newcommand{\ei}{\end{itemize}}

\newcommand{\Poincare}{Poincar\'{e}~}

\newcommand{\beq}{\begin{equation}}
\newcommand{\eeq}{\end{equation}}
\newcommand{\beqa}{\begin{eqnarray}}
\newcommand{\eeqa}{\end{eqnarray}}

\def\ltap{\ \raise.3ex\hbox{$<$\kern-.75em\lower1ex\hbox{$\sim$}}\ }
\def\gtap{\ \raise.3ex\hbox{$>$\kern-.75em\lower1ex\hbox{$\sim$}}\ }
\def\gl{\ \raise.5ex\hbox{$>$}\kern-.8em\lower.5ex\hbox{$<$}\ }
\def\roughly#1{\raise.3ex\hbox{$#1$\kern-.75em\lower1ex\hbox{$\sim$}}}

\begin{document}

		\thispagestyle{empty}
		\begin{flushright}
			March 2019\\
		\end{flushright}
		\vspace*{.2cm}
		\begin{center}			
			{\Large \bf An Algebraic Classification of Exceptional EFTs}
			\\
			\vspace*{.7cm} {\large Diederik Roest\footnote{\tt d.roest@rug.nl
				}, David Stefanyszyn\footnote{\tt
				d.stefanyszyn@rug.nl} and Pelle Werkman\footnote{\tt
				p.j.werkman@rug.nl}}

		\vspace{.3cm} {\em Van Swinderen Institute for Particle Physics and Gravity, University of Groningen, Nijenborgh 4, 9747 AG Groningen, The Netherlands}\\
\end{center}

		\vspace{1cm} 
We classify four-dimensional effective field theories (EFTs) with enhanced soft limits, which arise due to non-linearly realised symmetries on the Goldstone modes of such theories. We present an algorithm for deriving all possible algebras that can be non-linearly realised on a set of Goldstone modes with canonical propagators,  {linearly realised Poincar\'{e} symmetries} and interactions at weak coupling. We then perform a full classification of the cases with multiple scalars or multiple spin-$1/2$ fermions as the Goldstone modes. In each case there are only a small number of algebras consistent with field-dependent transformation rules, leading to the class of exceptional EFTs including the scalar sector of Dirac-Born-Infeld, Special Galileon and Volkov-Akulov theories. We also discuss the coupling of a $U(1)$ gauge vector to the exceptional scalar theories, showing that there is a Special Galileon version of the full Dirac-Born-Infeld theory. This paper is part I in a series of two papers, with the second providing an algebraic classification of supersymmetric theories.

	\vfill \setcounter{page}{0} \setcounter{footnote}{0}
	\newpage
	
	\tableofcontents
	

	\section{Introduction} \label{intro}
	
Effective field theories (EFTs) of Goldstone modes are particularly interesting quantum field theories where non-linearly realised symmetries yield special infra-red behaviour in scattering amplitudes. Recently, much effort has been devoted to classifying EFTs using both on-shell methods\cite{On-shell} (constructing amplitudes using minimal assumptions, see e.g. \cite{Direct,Gluons,SoftLimits1,SoftLimits2,SoftLimits3,Recursion,ScatteringEquations,SoftBootstrap} and references therein), and Lie-algebras (classifying the algebras that dictate the structure of these amplitudes \cite{LieAlgebraicScalar,Brauner2,LieAlgebraicVector}). Both approaches are completely insensitive to the Lagrangian basis for the Goldstone self-interactions and hence avoid Lagrangian redundancies associated with non-linear field redefinitions.

The simplest example of a Goldstone EFT with a non-linearly realised symmetry is that of a single, shift symmetric scalar field $\pi(x)$. While the Poincar\'{e} symmetries are linearly realised on the scalar, the shift symmetry ensures that a global $U(1)$ is non-linearly realised (see e.g. \cite{runaway} for a top-down approach to deriving this EFT). The shift symmetry is manifest in observables thanks to Adler's zero \cite{Adler}: when expanded around $p_{\text{soft}} = 0$, where $p_{\text{soft}}$ is the momentum attached to an external leg, all $\pi$-scattering amplitudes begin at linear order in this soft momentum.

A natural question to ask is if there are specific single scalar EFTs that have an enhanced soft limit, i.e.~where the leading order term in the soft amplitudes scales with \textit{non-linear} powers of $p_{\text{soft}}$. This was answered in a combination of works \cite{SoftLimits1,SoftLimits2,SoftLimits3} such that we now have a ``periodic table'' of such EFTs. It consists of Galileons \cite{Galileon} and the scalar Dirac-Born-Infeld\footnote{In this paper, we use \emph{scalar DBI} to refer to the non-linear realisation of a higher dimensional Poincar\'{e} symmetry describing the fluctuations of a probe brane in Minkowski space. We refer to the same theory coupled to a Born-Infeld vector as \emph{DBI}. In the literature, both of these theories are often referred to as simply "DBI".} (DBI) action \cite{BI,Dirac}  where the amplitudes begin at quadratic order in $p_{\text{soft}}$, and the Special Galileon \cite{SpecialGalileon} whose soft amplitudes begin at cubic order (see \cite{GeometryGal} for a detailed discussion on the Special Galileon coset space). The scalar DBI and Special Galileon EFTs are labelled {\it exceptional} since they have the maximal possible soft scaling\footnote{These exceptional theories   play a pivotal role in the double copy approach to amplitudes, see e.g.~\cite{ScatteringEquations, Double}.} for a given derivative power counting \cite{SoftLimits2}. 

These enhanced soft limits arise due to the presence of additional non-linearly realised symmetries beyond the scalar's shift symmetry which have explicit dependence on the space-time coordinates $x^{\mu}$. A soft limit of degree $\sigma$, where the leading order term in soft amplitudes scales as $p_{\text{soft}}^{\sigma}$, requires the existence of a symmetry which includes a field-independent term with $\sigma -1$ powers of $x^{\mu}$. Indeed the symmetry transformations which define Galileons and scalar DBI both have a linear dependence on the coordinates while the Special Galileon is invariant under a transformation rule with a quadratic dependence. At the level of transformation rules, the exceptional EFTs are those where at least one of the non-linear symmetries includes \textit{field-dependent} terms in its transformation rule in addition to the field-independent part\footnote{In this paper we always refer to the active form of the transformation rules where the coordinates do not transform, as opposed to the passive form where the coordinates can transform. If the reader is unfamiliar with this distinction we refer them to \cite{Olver}.}. This structure relates the coefficients of operators within the EFT which have differing mass dimensions. In particular, it relates the propagator to interactions. We note that this discussion applies to all Goldstone modes, not just scalars, and in this paper we will therefore use $\sigma$ to refer to the soft degree of any Lorentz representation.

Given that these symmetries have to form a consistent Lie-algebra with the Poincar\'{e} symmetries and the $U(1)$, a Lie-algebraic classification is also a very efficient way of examining the existence of these special EFTs and indeed those consisting of other irreducible representations of the Lorentz group. This approach is complementary to the amplitudes one and has been explored for scalars in \cite{LieAlgebraicScalar, Brauner2} and Abelian gauge vectors in \cite{LieAlgebraicVector}. Our primary aim in this paper is to outline an algorithm   to efficiently derive all possible algebras which can be non-linearly realised on a given set of Goldstone fields \textit{of any spin} under three assumptions:
\begin{itemize}
\item  Each algebra contains the four-dimensional Poincar\'{e} algebra (consisting of translations $P_{\mu}$ and Lorentz generators $M_{\mu\nu}$) which is linearly realised on the Goldstones\footnote{Our results therefore apply to fields in Minkowski space. We refer the reader to \cite{AdS} for recent work on non-linear symmetries in (anti)-de Sitter space.}. In addition, we allow for other linear internal symmetries, where appropriate, and non-linear generators. In the remainder of this paper we distinguish between linear and non-linear generators rather than unbroken and broken ones.
\item  At quadratic order in fields, each Goldstone has a canonical propagator. For non-linearly realised \emph{shift} symmetries, the kinetic term is the operator with the fewest powers of the field (assuming the absence of tadpoles). Therefore it must be invariant under the field-independent part of any transformation. Allowing for scaling symmetries and a dilaton, will give rise to the unique exception corresponding to non-linear realisations of conformal algebras, as discussed in section \ref{exceptional}.
\item The resulting low energy EFTs can be derived from the coset construction of non-linear realisations and \textit{inverse Higgs constraints} are necessary to eliminate any \textit{inessential} Goldstones.
\end{itemize}
This last   point might require some further explanation. For spontaneous breaking of space-time symmetries, Goldstone's theorem \cite{Goldstone} does not apply and there can be fewer Goldstone modes in the low energy EFT than non-linear generators\footnote{A very simple example which illustrates this nicely is that of superfluids \cite{Son} where a shift symmetric scalar is perturbed around a Lorentz breaking vacuum solution. There is only a single scalar fluctuation even though the generators of Lorentz boosts are spontaneously broken. See \cite{SymmetricSuperfluids} for a recent discussion on \textit{symmetric superfluids}. Our analysis does not capture superfluids since they do not linearly realise all Poincar\'{e} symmetries.}\cite{LowManohar}. In the coset construction of non-linear realisations \cite{Internal1, Internal2, Spacetime1, Spacetime2}, one introduces a Goldstone mode for every non-linear generator but for space-time symmetries one can then impose inverse Higgs constraints \cite{Spacetime2} to eliminate certain Goldstone modes while still non-linearly realising all broken symmetries. In this paper we call any Goldstone which can be eliminated by an inverse Higgs constraint \textit{inessential}\footnote{These fields are only inessential in terms of non-linearly realising the symmetries at low energies but may well indeed be an important part of any (partial) UV completion \cite{UV-InverseHiggs}.} as opposed to an \textit{essential} one which cannot. In terms of enhanced soft limits, inverse Higgs constraints are a crucial ingredient: as we discuss in detail in section \ref{Spacetime}, the existence of explicit powers of $x^{\mu}$ in symmetry transformations, which is necessary to realise enhanced soft limits, \textit{requires} the existence of inverse Higgs constraints.

There are two different types of algebras which are of interest. In the first case the only non-vanishing commutators are those required by the existence of inverse Higgs constraints and those which define the Lorentz representation of the generators. Here all transformation rules on the essential Goldstones are field-independent and correspond to extended shift symmetries \cite{HinterbichlerExtended}, i.e. polynomials in the space-time coordinates. These algebras always exist: once the conditions for inverse Higgs constraints have been met, these algebras satisfy all Jacobi identities automatically. In this case the enhancement of soft limits is trivial unless the operators are Wess-Zumino terms which have fewer derivatives per field than the strictly invariant operators (e.g.~the leading order interactions of the scalar Galileon are very well known examples of Wess-Zumino terms \cite{GalWess} with non-trivial soft behaviour). The second case is defined by having at least one non-vanishing commutator between two non-linear generators. This leads to field-dependent transformation rules for the Goldstones and exceptional EFTs. We are mainly interested in algebras of this type (as suggested by the title of this paper) since these algebras are more difficult to construct and lead to more restricted dynamics.

To classify exceptional symmetry algebras, our algorithm essentially boils down to three separate steps. These will be explained in detail in section \ref{Spacetime} but let us summarise them here:
\begin{itemize}
\item \textit{Step I}: Construct the ``inverse Higgs trees''. This involves constraining the presence of non-linear generators on the right-hand side of commutators between translations $P_{\mu}$ and non-linear generators. This is done by requiring the existence of inverse Higgs constraints to eliminate inessential Goldstones, and satisfying Jacobi identities involving two copies of translations and one non-linear generator, up to the presence of linear generators. If there are multiple essential Goldstones then there are multiple, decoupled inverse Higgs trees and the result of this step is that all non-linear generators in a given tree must live in the Taylor expansion of the corresponding essential Goldstone. 

For example, if we have a single essential scalar Goldstone we can only add fully symmetric generators in the non-linearly realised algebra. Similarly, if we have a single essential vector Goldstone all non-linear generators can have at most one pair of anti-symmetric indices.
\item \textit{Step II}: Demand invariance of the canonical kinetic term for each Goldstone under the field-independent part of every transformation rule. The field-independent part of the transformation rules are fixed by step I. If a canonical kinetic term is to exist in the resulting EFT, this is a necessary requirement since the canonical kinetic term is the operator with the fewest powers of the field (unless the EFT includes the dilaton which turns out to be the unique exception). This condition constrains the Lorentz structure of the allowed non-linear generators and provides a very powerful constraint on the non-linear algebras.  In addition to the kinetic terms, the non-linearly realised symmetries allow for interactions between Goldstone modes which should be considered as perturbative corrections \cite{Vainshtein}.

\item \textit{Step III}: Place further constraints on the algebra by satisfying the remaining Jacobi identities. Generally speaking, the best order to calculate them in is the following: first constrain the presence of linear generators on the right-hand side of commutators between translations and non-linear generators by considering Jacobi identities involving two copies of translations. Then consider Jacobi identities involving one copy of translations and two non-linear generators and finally those involving three non-linear generators. All Jacobi identities involving the generators of Lorentz are automatically satisfied as long as the commutators of the algebra involve only Lorentzian multiplets which is always the case in this paper.
\end{itemize}
We then apply our algorithm to a number of physically relevant cases. In each case, other than non-linear realisations of conformal algebras, the existence of canonical propagators ensures that the number of inverse Higgs constraints we must impose to reduce to only the essential Goldstones is equivalent to $\sigma -1$ for each inverse Higgs tree, as shown in section \ref{stepII}. We then derive all possible algebras which can be non-linearly realised and our results can be summarised as follows:
\begin{itemize}
\item Section \ref{realscalar}: In the case of a single scalar, field-dependent transformation rules can only arise in the presence of at most two inverse Higgs constraints. If one is required to impose three or more inverse Higgs constraints to arrive at the single scalar EFT, all transformation rules must be field-independent i.e. reduce to extended shift symmetries. Since the number of inverse Higgs constraints is equal to $\sigma -1$, our results tell us that single scalar EFTs with $\sigma >3$ cannot be invariant under field-dependent symmetry transformations. This implies that there are no exceptional real scalar EFTs with $\sigma >3$, as found in \cite{SoftLimits2} via on-shell methods. Therefore, the only exceptional EFTs are those of the known scalar DBI and the Special Galileon. There is also the non-linear realisation of the four-dimension conformal algebra resulting in the dilaton EFT describing the dynamics of a flat probe brane in five dimensional AdS space. However, here there is no Adler's zero and therefore no enhanced soft limit either.

\item Section \ref{multiscalar}: In the case of multiple scalars, the exceptional EFTs are very similar to the single-field case. When no Goldstone satisfies more than one inverse Higgs constraint, the most general exceptional EFT is built from three ingredients: firstly, a multi-scalar DBI sector. These form a non-linear realisation of higher dimensional Poincar\'{e} algebras, with the dimension of the algebra fixed by the number of scalars: each scalar parametrisations an extra dimension \cite{multiDBI}. Secondly, the DBI scalars can couple to a number of Galileons, which represent particular contractions of Poincar\'e algebras. Lastly, one can couple these two sectors to some internal coset space $G' / H'$. Again the presence of a single inverse Higgs constraint tells us that this EFT has at most $\sigma = 2$ soft behaviour on each of the scalars.

If we have to impose two or more inverse Higgs constraints on any Goldstone, then one Goldstone can be a Special Galileon. However, the others must have empty inverse Higgs trees, i.e. they are the Goldstone modes of broken internal symmetry groups. This proves that there is no multi-scalar version of the Special Galileon\footnote{We thank James Bonifacio for pointing out that this might be very interestingly related to the fact one cannot couple two massless gravitons to each other.}, as was suggested by the amplitude results of \cite{SoftBootstrap}. In the presence of additional inverse Higgs constraints, all symmetries must reduce to extended shift symmetries. The only other possibilities in addition to the above include the dilaton coupled to axions and correpond to non-linear realisations of higher dimensional conformal algebras.  

\item Section \ref{fermion}: For $N$ fermions, field-dependent transformation rules are incompatible with the existence of inverse Higgs constraints. If any inverse Higgs constraints need to be imposed, only extended shift symmetries are possible. Examples include a fermionic shift symmetry \cite{VA-SmallFieldLimit} and a shift linear in the coordinates, which would be the fermionic generalisation of the scalar Galileon. The generator of this transformation rule has spin-$3/2$. In the absence of inverse Higgs constraints the exceptional EFTs correspond to non-linear realisations of extended supersymmetry i.e. Volkov-Akulov \cite{VA,IvanovKapustnikov,IvanovKapustnikov2} and its multi-field extension (which can moreover be coupled to an arbitrary number of shift symmetric fermions in comparison to the scalar DBI and Special Galileon cases mentioned above). The soft behaviour in this case is restricted to $\sigma =1$ since there are no inverse Higgs constraints. We refer the reader to \cite{FermionSoft1,FermionSoft2} for discussions on the soft behaviour of Volkov-Akulov and \cite{BrandoOther} for phenomenological applications of the multi-field extension.

\item Section \ref{vectoressential}: One can couple a gauge vector to the scalar Goldstone of an exceptional EFT as long as it transforms appropriately under the generators in the scalar's inverse Higgs tree. The gauge vector is not a Goldstone mode in these theories but rather transforms as a matter field. Nevertheless, this restricts the couplings between the scalar Goldstone and the gauge vector significantly, leading to either the Dirac-Born-Infeld theory with a scalar and a vector, or a Special Galileon analogon thereof. We expect that the latter explains the scalar-vector couplings found in \cite{SoftBootstrap}. It would be very interesting to investigate any connections between this symmetry, with its scalar-vector couplings, and decoupling limits of massive gravity \cite{MG,MG2}.
\end{itemize}

Before moving to the main body of the paper, we would like to emphasise that our results are purely based on the existence of Lie-algebras and the above assumptions. We do not have to assume anything about the structure of the leading order Goldstone interactions. This algebraic method therefore offers definitive answers as to whether symmetries can exist which is potentially an advantage over the amplitudes approach. Furthermore, the techniques outlined here enable us to perform full classifications for any number of scalars and any number of fermions which we imagine would require much more work to arrive at the same results using soft amplitudes.

Our results have clear implications for model building in both cosmology and particle physics. Indeed, non-linear symmetries can play a vital role in both and our results in the context of multi-scalars and multi-fermions present all possibilities. We end this paper with our concluding remarks and briefly introduce the second in this series of papers where we replace the linear Poincar\'{e} symmetries with those of $\mathcal{N}=1$ super-Poincar\'{e}. We also speculate on a number of directions for future work.


\section{Goldstone modes: lost in translations} \label{Spacetime}
In this section we outline the three steps in our algorithm. The first step is to incorporate the necessary inverse Higgs constraints at the level of the algebra, while also satisfying Jacobi identities which involve two copies of translations. This results in \textit{inverse Higgs trees}. The second step is to demand the presence of canonical propagators in the resulting EFTs, which restricts the Lorentz representation of allowed non-linear generators in a very powerful way. This is possible since in the absence of the dilaton and tadpoles, the kinetic terms are the operators with the fewest powers of the field. After the first two steps the algebra has been restricted as much as possible in a model-independent manner. The third step is to then impose the final constraints from Jacobi identities.
\subsection{Step I: inverse Higgs trees} \label{stepI}
Consider a theory with symmetry group $G$. Whenever this theory has a vacuum that breaks this symmetry group down to a subgroup $H$, the generators $G_i$ which live in the coset space $G / H$ will induce fluctuations around the vacuum field configuration $\vert 0 \rangle$:
\begin{equation}
\phi^{i}(x) G_{i} \vert 0 \rangle \, .
\end{equation}
When the non-linear generators are internal, Goldstone's theorem tells us that each generator induces an independent massless fluctuation. In other words, the number of massless modes $\phi^i$ equals the number of non-linear generators $G_i$, with $i=0, \ldots, n$ (here we are suppressing any Lorentz structure and indeed the following discussion applies for Goldstone modes of any Lorentz representation). However, when space-time symmetries are spontaneously broken, the Goldstone modes generated by $G_i$ may not all be linearly independent. This requires \cite{LowManohar}
\begin{align}\label{eq:LowManohar}
\phi^{i}(x) G_{i} \vert 0 \rangle = 0 \,,
\end{align} 
to have non-trivial solutions. These solutions relate the Goldstone modes and thus distinguish between essential Goldstones, forming an integral part of the Goldstone EFT, and inessential ones that can be solved for in terms of the essentials. In order to see this relation at lowest order explicitly, one can act on \eqref{eq:LowManohar} with a space-time derivative yielding
\begin{align}
0 &= \partial_{\mu} (\phi^{i} G_{i}) \vert 0 \rangle =  (\partial_{\mu} \phi^i - \phi^{j} f_{\mu j}{}^{i} ) G_i \vert 0 \rangle, \label{IHLow}
\end{align}
where we have allowed for the most general commutator between translations $P_\mu = - \partial_\mu$ and non-linear generators, 
\begin{align}
[P_{\mu},G_{i}] = i f_{\mu i}{}^{j}G_{j} + {\rm linear ~ generators} \,, \label{IHCondition}
\end{align}
where the latter term can contain the linear generators (including $P_\mu$, $M_{\mu \nu}$). There is always at least one generator $G_0$ which satisfies $[P_\mu, G_0] = {\rm linear ~ generators}$. Such a generator must correspond to an essential Goldstone mode, so we dub it an \emph{essential generator}.

One can satisfy the condition \eqref{IHLow} by imposing
\begin{equation}
\partial_{\mu} \phi^i - \phi^j f_{\mu j}{}^{i} + \mathcal{O}(\phi^{2})= 0 \,,
\end{equation}
where the higher-order terms will be determined by different commutation relations that we will discuss later on. 
For $f_{\mu j}{}^{i} \neq 0$ (i.e. when $[P_\mu, G_j] \supset G_i$), this tells us that one Goldstone mode and the space-time derivative of another one are linearly dependent, even though the corresponding generators are linearly independent. This indicates that we can reduce the number of Goldstone modes.  

We now focus on a particular essential Goldstone $\phi^0$. The inessential Goldstones related to $\phi^0$ are given by 
\begin{align}\label{eq:firstorder}
\partial_\mu \phi^0 = f_{\mu i}{}^{0} \phi^i  + \mathcal{O}(\phi^{2}) \,.
\end{align}
Note that we have not required $G_i$ to be irreducible Lorentz representations, and indeed a generic covariant derivative of $\phi^{0}$ will include a number of irreps which we have collectively denoted as $\phi^i$ (and are implicitly summed over). Acting with a second derivative on \eqref{IHLow} yields additional relations between the Goldstone modes 
\begin{align}
\partial_\mu \partial_\nu \phi^0 =   f_{\mu i}{}^{j} f_{\nu j}{}^{0} \phi^i  + \mathcal{O}(\phi^{2})  \,. \label{second-order}
\end{align}
Again, this relation may be imposed to project out inessential Goldstone modes in terms of derivatives operators acting on $\phi^0$. The linear combination of generators appearing on the right-hand side of \eqref{second-order} has a commutator with translations that includes the right-hand side of \eqref{eq:firstorder}, i.e. $f_{\mu i}{}^{j} f_{\nu j}{}^{0} \neq 0$. 

\begin{wrapfigure}[15]{R}{7cm}
	\vspace{-0.5cm}
	\begin{tikzpicture}
	\node at (0, 1){\textbf{Representations}};
	\draw [dashed, red, thin] (0+0.2,0-0.2) -- (1.5 - 0.2,-1.5 + 0.2);
	\draw [dashed, blue, thin] (0-0.2,0-0.2) -- (-1.5 + 0.2,-1.5 + 0.2);
	\draw [dashed, red, thin] (0+0.03,0-0.2) -- (0 + 0.03,-1.5 + 0.2);
	\draw [dashed, blue, thin] (0-0.03,0-0.2) -- (0 - 0.03,-1.5 + 0.2);
	\node at (0,0){$(s)$};
	\draw [dashed, red, thin](-1.5 - 0.2, -1.5 - 0.2) -- (-3 +0.2, -3 + 0.2);
	\draw [dashed, red, thin](-1.5+0.2,-1.5-0.2) -- (0-0.2,-3+0.2);
	\draw [dashed, red, thin](0 - 0.16, -1.5 - 0.2) -- (-1.5 +0.24, -3 + 0.2);
	\draw [dashed, red, thin](0+0.24,-1.5-0.2) -- (1.5-0.16,-3+0.2);
	\draw [dashed, blue, thin](0 - 0.24, -1.5 - 0.2) -- (-1.5 +0.16, -3 + 0.2);
	\draw [dashed, red, thin](0+0.16,-1.5-0.2) -- (1.5-0.24,-3+0.2);
	\draw [dashed, red, thin](1.5 + 0.2, -1.5 - 0.2) -- (3 -0.2, -3 + 0.2);
	\draw [dashed, red, thin](1.5-0.2,-1.5-0.2) -- (0+0.2,-3+0.2);
	\draw [dashed, blue, thin] (-1.5+0.05,-1.5-0.2) -- (-1.5 + 0.05,-3 + 0.2);
	\draw [dashed, red, thin] (-1.5-0.05,-1.5-0.2) -- (-1.5 - 0.05,-3 + 0.2);
	\draw [dashed, red, thin] (1.5+0.05,-1.5-0.2) -- (1.5 + 0.05,-3 + 0.2);
	\draw [dashed, red, thin] (1.5-0.05,-1.5-0.2) -- (1.5 - 0.05,-3 + 0.2);
	\draw [dashed, red, thin] (0+0.05,-1.5-0.2) -- (0 + 0.05,-3 + 0.2);
	\draw [dashed, red, thin] (0-0.05,-1.5-0.2) -- (0 - 0.05,-3 + 0.2);
	\draw [dashed, red, thin] (0+0.09,-1.5-0.2) -- (0 + 0.09,-3 + 0.2);
	\draw [dashed, red, thin] (0-0.09,-1.5-0.2) -- (0 - 0.09,-3 + 0.2);
	\node at (-1.5,-1.5) {$(s-1)$};
	\node at (0,-1.5) {$(s)$};
	\node at (1.5,-1.5) {$(s+1)$};
	\node at (-3,-3) {$(s-2)$};
	\node at (-1.5,-3) {$(s-1)$};
	\node at (0,-3) {$(s)$};
	\node at (1.5,-3) {$(s+1)$};
	\node at (3,-3) {$(s+2)$};
	\end{tikzpicture}
	\caption{\it The possible non-linear symmetries that can be realised on a spin-$s$ Lorentz representation and their links via space-time translations.}
	\label{fig:scalartree1}
\end{wrapfigure}
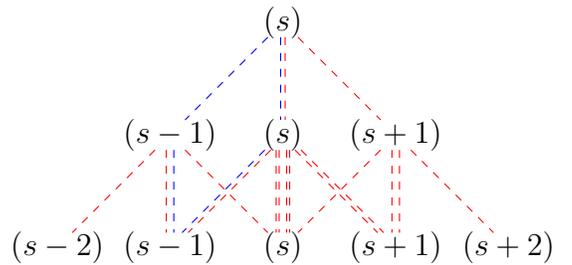
The general procedure should now be clear: one starts off with an essential Goldstone mode and applies a space-time derivative $n$ times to \eqref{IHLow}. This yields an equation involving $n$ factors of structure constants that may be imposed as a constraint to project out an \emph{$n^{\text{th}}$-order} inessential Goldstone mode. The generator corresponding to an $n^{\text{th}}$-order inessential must contain an $(n-1)^{\text{th}}$-order inessential in its commutator with $P_{\mu}$ in order for such a constraint to be possible. We represent this algebraic structure, the \emph{inverse Higgs tree}, in figure \ref{fig:scalartree1}. The lines connecting generators indicate that the lower one contains the upper one in its commutation relation with translations. 

Starting the inverse Higgs tree with a spin $s$ essential Goldstone allows for spins $s-1$ to $s+1$ at first order, with degeneracies $(1,2,1)$ from left to right. Similarly, at second order one a priori has spins ranging from $s-2$ to $s+2$ with degeneracies $(1,4,6,4,1)$. However, there will be consistency conditions on the algebra that reduce this degeneracy, as we will argue later in this section. At this point, the lines in the tree do not imply that other non-linear generators are not present in the commutation relations with translations. A generator at second order with spin $(s - 1)$ may have a direct connection with the essential spin $s$, seemingly spoiling the ordering of the algebra in terms of levels in the inverse Higgs tree. Additionally, when there are several essential generators and Goldstones, a particular generator may a have a commutation relation with $P_\mu$ that contains generators appearing in more than one tree i.e. associated with more than one essential Goldstone, making the structure of inverse Higgs relations unclear. 

However, by choosing a convenient basis for the generators, we can show that the structure of inverse Higgs relations is unique. The non-linear generators in our algebra coincide with a particular choice of paths in the inverse Higgs trees of  the essential generators we include. That is, we assign linearly independent generators to each of a set of nodes in a path connected to the essentials. Then, the commutation relation of each generator with translations contains at least the generator that comes before it in the inverse Higgs tree. We can then label each generator $G^{(i,n,p)}_{\mu_1 \ldots \mu_n}$ according to its Lorentz representation, the tree $i$ (with one tree for each essential Goldstone), node $n$ and path $p$ it is assigned to. The commutation relations satisfy
\begin{equation}\label{eq:weakordering}
[P, G^{(i,n,p)}] = i G^{(i, n-1, p)} + \ldots \, ,
\end{equation}  
where the ellipses indicate non-linear generators besides $G^{(i,n-1,p)}$ and linear generators. Additionally, we assume that none of the commutation relations between non-linear generators and translations contain one of the end points of the chosen paths: $[P, G] \not\supset G_{end}$ for any $G$ and $G_{end}$. In general, this sort of commutation relation is possible\footnote{A simple example of this set-up would be a scalar and a two-form essential Goldstone which both have the same inessential vector at level one in their respective trees.}. 

However, in the cases we consider in this paper (scalar, spin-$\frac{1}{2}$, vector, and combined scalar and vector essential generators) such a commutation relation can never occur. In the case where there is a single essential generator, we can show in generality that this scenario is impossible: if a generator $\tilde{G}$ in a particular tree commuted into an end-point of that tree under translations, that would amount to introducing a new generator for the node below the end-point and identifying it with $\tilde{G}$. Such an identification implies that acting sequentially with translations yields repeating patterns, rather than ever ending on an essential generator. 

With this restriction, it is clear that we can remove non-linear generators other than $G^{(i, n-1, p)}$ from the right-hand side of \eqref{eq:weakordering}. Since all $[P, G]$ cannot contain end-point generators, there is a linear combination of generators independent from $G^{(i, n,p)}$ (but of the same Lorentz representation) that yields the combination of generators appearing in the ellipses. We can simply subtract that linear combination from $G^{(i, n, p)}$ by a change of basis. We therefore obtain
\begin{equation}\label{eq:strongordering}
[P, G^{(i,n,p)}] = i G^{(i, n-1, p)} + {\rm linear ~ generators} \,.
\end{equation}  
Therefore, choosing a particular set of paths in the algebraic structure displayed in figure \ref{fig:scalartree1} fixes the inverse Higgs relations one can impose. In particular, when we have several essential generators, the inverse Higgs trees decouple as far as the commutation relations with translations are concerned. This final point is an important part of our ability to perform exhaustive classifications for multi-scalars and multi-fermions.

Let us return to equation \eqref{second-order} and see what sort of consistency conditions it imposes on our algebra. Clearly, the left-hand side is symmetric in $(\mu \leftrightarrow \nu)$. This imposes $f_{[ \mu 2}{}^{1} f_{\nu] 1}{}^0  =0$. The same condition may be derived from imposing Jacobi identities on the structure constants appearing in \eqref{second-order}. The Jacobi identity involving two translations and a non-linear generator at second order, $(P, P, G_2)$, relates the different routes from the second order Goldstones to the essential. Picking e.g.~a specific second order generator with $s-1$, the two routes from the essential towards it (via spin $s$ and $s-1$ at first order, forming a parallelogram) are related due to the symmetry, as indicated in blue in figure \ref{fig:scalartree1}. In total, these reduce the independent generators at second order to $(1,2,4,2,1)$, identical to a Taylor expansion to quadratic order of a spin $s$ representation.

The observation \eqref{second-order} restricts the possible space-time symmetries which can be realised by $\phi^{0}$. For example, if the essential Goldstone is a scalar the non-linearly realised algebra can only contain symmetric generators at any order, and no anti-symmetric ones (as also found after studying Jacobi identities  \cite{LieAlgebraicScalar}). A similar argument applies to an essential vector Goldstone which cannot realise an algebra with $p$-form generators with $p > 2$ \cite{LieAlgebraicVector}. Similarly, leaving out generators will have implications for generators higher up in the tree structure: due to the Jacobi identities, leaving out the first-order generators with spin $s-1$ will also imply the absence of the second-order generators with spin $s-2$ and $s-1$.

\begin{wrapfigure}[18]{R}{5cm}
	\begin{tikzpicture} 
	\node at (2, 1.0){\textbf{Scalar tree}};
	\draw [dashed, red, thin] (0+0.2,0-0.2) -- (1 - 0.2,-1 + 0.2);
	\node at (0,0){$\bullet$};
	\draw [dashed, red, thin](1 + 0.2, -1 - 0.2) -- (2 -0.2, -2 + 0.2);
	\draw [dashed, red, thin](1-0.2,-1-0.2) -- (0+0.2,-2+0.2);
	\node at (1,-1) {\tiny $\yng(1)$};
	\node at (0,-2) {$\bullet$};
	\node at (2,-2) {\tiny $\yng(2)$};
	\draw [dashed, red, thin](0 + 0.2, -2 - 0.2) -- (1 - 0.2, -3 + 0.2);
	\draw [dashed, red, thin](2 + 0.2, -2 - 0.2) -- (3 - 0.2, -3 + 0.2);
	\draw [dashed, red, thin](2 - 0.2, -2 - 0.2) -- (1 + 0.2, -3 + 0.2);
	\node at (1,-3) {\tiny $\yng(1)$};
	\node at (3, -3) {\tiny $\yng(3)$};
	\draw [dashed, red, thin](1 -0.2, -3 - 0.2) -- (0 + 0.2, -4 + 0.2);
	\draw [dashed, red, thin](1 +0.2, -3 -0.2) -- (2 -0.2, -4 + 0.2);
	\draw [dashed, red, thin](3 - 0.2, -3 - 0.2) -- (2 + 0.2, -4 + 0.2);
	\draw [dashed, red, thin](3 + 0.2,  -3 - 0.2) -- (4 - 0.2, -4 + 0.2); 
	\node at (0,-4){$\bullet$};
	\node at (2, -4){\tiny $\yng(2)$};
	\node at (4, -4){\tiny $\yng(4)$};
	\end{tikzpicture}
	\caption{\it The non-linear symmetries that can be realised on a scalar, and their space-time relations.}
	\label{fig:scalartree}
\end{wrapfigure}
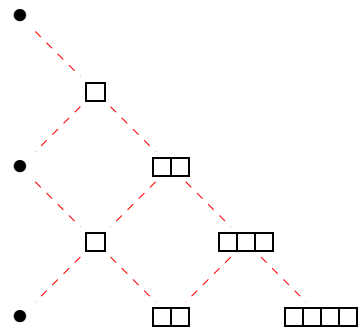

As an example, the essential scalar inverse Higgs tree is displayed in figure \ref{fig:scalartree}. At first order in the tree we have an inessential vector while at second order we have a scalar and a symmetric traceless generator. The higher order structure follows straightforwardly with all generators appearing in the scalar's Taylor expansion. In the figure each generator is an irrep and when a generator is connected by lines to several above it, all these connections must be there simultaneously for consistency with Jacobi identities.

The above general discussion follows from the first part of the commutator between translations and Goldstones \eqref{IHCondition}. It does not restrict the linear generators that can appear on the right-hand side of this commutator. Moreover, it does not involve the commutator between a pair of non-linear generators. These parts of the algebra will become important, however, at higher order in fields. These follow from the coset construction of non-linear realisations, both for internal \cite{Internal1, Internal2} and space-time symmetries\footnote{In contrast to the internal case (with only scalar generators) which has coset universality, for space-time symmetries there is no proof that all non-linear realisations are equivalent to a particular coset construction.} \cite{Spacetime1, Spacetime2}. It can be seen as a generalisation of the covariant derivatives \eqref{IHLow} with higher-order, field-dependent terms. Setting these non-linear conditions to zero is referred to as imposing inverse Higgs constraints\footnote{See \cite{KRS, McArthur} for more detailed discussions on the necessary requirements for the existence of inverse Higgs constraints.} \cite{Spacetime1, Spacetime2}. 

The ordering of inessential generators manifests itself in the transformation rules of the essential Goldstone $\phi^0$. Starting with the lowest level generator $G_0$, it acts on the essential Goldstone as
\begin{align}
\delta_{0} \phi^0 = c_0 + \ldots \,,
\end{align}
where $c_0$ is a constant, again with any Lorentz structure suppressed, while $\ldots$ represents field-dependent terms following from the coset construction. Due to \eqref{second-order} and its generalisations, higher order generators act as monomial shifts in the coordinates plus field-dependent terms,
\begin{align} \label{GenShifts}
\delta_{n} \phi^0 = c_{n} x^{n} + \ldots \,,
\end{align}
where $c_{n}$ are constants in the appropriate Lorentz representations. In the absence of field-dependence, these symmetries are referred to as extended shift symmetries of order $n$ \cite{HinterbichlerExtended}. This again highlights the relation to the Taylor expansion of the essential Goldstone mode. Furthermore, as explained in the introduction, this implies that the addition of higher-order generators makes the scattering amplitudes for the essential Goldstone softer and softer since the soft degree increases linearly in $n$. 

While the invariant operators of the order $n$ extended shift symmetries require at least $n + 1$ derivatives per field, this is not true in general for their Wess-Zumino terms. Moreover, when the transformation law \eqref{GenShifts} contains field-dependent terms, this relation between the order in the inverse Higgs tree and the number of derivatives can be violated. This is what gives the exceptional EFTs their enhanced soft limit at a given power counting. As we will see later in the paper, the leading operators of the higher-dimensional Anti-de Sitter algebras are a third exception to this counting of derivatives, even though they are invariants (not Wess-Zumino terms) without enhanced soft limits. It is therefore natural to wonder what other exceptions to the naive counting of derivatives can occur at higher orders in the inverse Higgs trees. In the following sections, we will carry out an exhaustive classification to arbitrary finite order in the inverse Higgs tree for different choices of essential Goldstones.

\subsection{Step II:  canonical propagators} \label{stepII}
The existence of a canonical propagator for the essential Goldstones further restricts the Lorentz representation of the allowed non-linear generators. For bosons, the second order differential operator \eqref{second-order} will contain a projection that also appears in the kinetic term of the essential Goldstone. This particular generator, with a specific transformation $\delta_{2} \phi^0 = c_2 xx$, will therefore not leave the kinetic term invariant. Since the latter is the operator with the fewest powers of the field (at least in the absence of tadpoles and the dilaton) this is not a viable part of any transformation rule, even if it is augmented with other terms. This rules out the possibility of adding this particular second order generator in any non-linearly realised algebra. We will illustrate how this can be a powerful constraint on the algebra in the context of the examples we consider in section \ref{exceptional} i.e. scalars, fermions and vectors. 

First consider an essential scalar Goldstone denoted $\pi$. We know from our step I discussion that all non-linear generators must be fully symmetric, since they must live in the scalar's Taylor expansion. The field-independent part of the transformation rule generated by $n^{\text{th}}$ order non-linear generators is therefore
\begin{align}
\delta_{n}\pi = s_{\mu_{1}, \ldots, \mu_{n}}x^{\mu_{1}} \ldots x^{\mu_{n}}
\end{align} 
where for now $s_{\mu_{1},\ldots, \mu_{n}}$ are not Lorentzian irreps since they include traces. Now invariance of the kinetic operator $\pi \Box \pi$ up to a total derivative requires the symmetry parameter (and the corresponding generator) to be traceless. Of course, at each order in $n$ there is only a single generator with this property, which has spin $n$. Therefore, the inverse Higgs tree collapses onto a single line: the top diagonal in figure \ref{fig:scalartree}. As a consequence, since a soft degree of order $\sigma$ requires the presence of a symmetry transformation with $\sigma -1$ powers of $x^{\mu}$, the number of inverse Higgs constraints required to reduce to the essential Goldstone mode is equivalent to $\sigma -1$. This analysis carries over to any number of essential scalars given that, as discussed in section \ref{stepI}, the inverse Higgs trees for different essential Goldstones decouple: in each tree we can only add fully symmetric and traceless generators at each order.

Next consider an essential spin-$1/2$ fermion $\lambda_{\alpha}$. Since we are specialising to four space-time dimensions in this paper, in the following we will employ $SU(2) \times SU(2)$ notation for \textit{all} space-time indices using $(\sigma^{\mu})_{\alpha \dot{\alpha}}$ to map between $SU(2) \times SU(2)$ and $SO(1,3)$ e.g. for translations we have $P_{\alpha \dot{\alpha}} = (\sigma^{\mu})_{\alpha \dot{\alpha}} P_{\mu}$. In this notation the Weyl kinetic term for the fermion is simply $\lambda_{\alpha} \partial^{\alpha \dot{\alpha}} \bar{\lambda}_{\dot{\alpha}}$. Again from step I we know that all generators in the fermion's inverse Higgs tree must live in the Taylor expansion of the fermion i.e. the field-independent part of the transformation rules at order $n$ are
\begin{align}
\delta_{n} \lambda_{\alpha} = t_{\alpha \beta_{1},\ldots, \beta_{n} \dot{\beta}_{1},\ldots, \dot{\beta}_{n}}x^{\beta_{1}\dot{\beta}_{1}}\ldots x^{\beta_{n}\dot{\beta}_{n}}
\end{align}
where again the parameters $ t_{\alpha \beta_{1},\ldots, \beta_{n} \dot{\beta}_{1},\ldots, \dot{\beta}_{n}}$ are not irreps but are symmetric in exchange of $\beta_{i}$ and $\beta_{j}$ if we also exchange $\dot{\beta}_{i}$ and $\dot{\beta}_{j}$. Demanding that the Weyl kinetic term is invariant under these transformations up to a total derivative requires the parameters to be fully symmetric in the sets $(\alpha,\beta_{1},\ldots,\beta_{n})$ and $(\dot{\beta}_{1},\ldots \dot{\beta}_{n})$. Therefore, in comparison to the essential scalar case, at each order in the inverse Higgs tree there is a single irrep which can be included which in this fermionic case has spin\footnote{We remind the reader that in $SU(2) \times SU(2)$ notation, one can only take traces with the anti-symmetric tensors $\epsilon_{\alpha \beta}$, $\epsilon_{\dot{\alpha}\dot{\beta}}$ so fully symmetric generators are irreps.} $\frac{1}{2}(1+n)$. It follows that again the number of inverse Higgs constraints is equal to $\sigma-1$. The extension to multiple fermions is trivial, since there the inverse Higgs trees again decouple so we can only include a single generator at each order in each tree.

Finally consider an essential $U(1)$ gauge vector $A_{\mu}$ which follows very similarly to the previous discussions. Again all field-independent parts of the transformation rules must live in the Taylor expansion of the essential vector i.e. they must take the form 
\begin{align} \label{vectorsymms}
\delta_{n}A_{\mu} = u_{\mu \nu_{1}, \ldots, \nu_{n}}x^{\nu_{1}} \ldots x^{\nu_{n}}
\end{align} 
where again $u_{\mu \nu_{1},\ldots, \nu_{n}}$ are not Lorentz irreps. However, they must be symmetric in $(\nu_{1},\ldots \nu_{n})$. There are therefore two distinct possibilities at each order in $n$: either $u_{\mu \nu_{1},\ldots, \nu_{n}}$ is fully symmetric in which case \eqref{vectorsymms} generates gauge transformations or $u_{\mu \nu_{1},\ldots, \nu_{n}}$ has one pair of anti-symmetric indices. This last possibility is the interesting one, since it can in principle constrain vector EFTs further than simply requiring gauge symmetry. If symmetries of this type are to leave the $U(1)$ kinetic term $F_{\mu\nu}F^{\mu\nu}$ (where $F_{\mu\nu} = \partial_{\mu}A_{\nu} - \partial_{\nu}A_{\mu}$) invariant up to a total derivative, only the traceless part of these generators can be included in any non-linearly realised algebra. At each order in $n$ we therefore have gauge symmetries plus one extra non-linear generator.

It should be noted that the above discussion can be (but is not necessarily) modified in the presence of several Goldstones. For instance, when classifying the EFTs of a number of scalar Goldstones that all have a constant shift symmetry, the lowest order terms in a derivative expansion are the two-derivative kinetic terms, and hence the above discussion applies. We will investigate this in detail in section \ref{multiscalar}. However, if one also has a dilaton-like Goldstone that has a non-linear \emph{scaling}  symmetry, the lowest order terms will be potential terms and hence the above discussion will be modified: the relevant representations are not restricted to the diagonal of Figure \ref{fig:scalartree}. Interestingly, there is a unique case of this type that arises in the EFTs based on higher-dimensional Anti-de Sitter algebras. This will also be discussed at the appropriate point in the next chapter.

\subsection{Step III: towards exceptional EFTs}

Step I and step II have now severely constrained the allowed algebras that can be non-linearly realised. We have fixed the right-hand side of all commutators between non-linear generators and translations, up to the presence of linear generators. We have also eliminated many non-linear generators at each order by demanding that the Goldstone EFT contains canonical propagators for each essential Goldstone. We have seen that the latter is very powerful for scalars, fermions and vectors and will enable us to perform exhaustive classifications in the following section. Step III is now very simple and requires us to satisfy the remaining Jacobi identities in the following order: first constrain the presence of linear generators on the right-hand side of commutators between translations and non-linear generators by considering Jacobi identities involving two copies of translations, then consider Jacobi identities involving one copy of translations and two non-linear generators and finally those involving three non-linear generators.

The aim of the game is to then find examples where each of these steps allows for non-linear algebras with at least one non-zero commutator between non-linear generators. These lead to field-dependent transformation rules for the Goldstones and exceptional EFTs.

\section{Exceptional EFTs} \label{exceptional}

In this section we illustrate our algorithm using a number of examples where we can exhaustively classify all algebras that can be non-linearly realised. We will again employ $SU(2) \times SU(2)$ indices, since we are working in four space-time dimensions. We use the following convention for commutators between a tensor $T_{\alpha_{1},\ldots \alpha_{n} \dot{\alpha}_{1},\ldots \dot{\alpha}_{m}}$ and the Lorentz generators $M_{\beta \gamma}$, $\bar{M}_{\dot{\beta}\dot{\gamma}}$
\begin{align}
[T_{\alpha_{1}\ldots \alpha_{n} \dot{\alpha}_{1}\ldots \dot{\alpha}_{m}} , M_{\beta \gamma}] &= 2 \, n! \, i\epsilon_{\alpha_{1} (\beta} T_{\gamma) \alpha_{2} \ldots \alpha_{n} \dot{\alpha}_{1} \ldots \dot{\alpha}_{m}} \,, \notag \\
[T_{\alpha_{1}\ldots \alpha_{n} \dot{\alpha}_{1} \ldots \dot{\alpha}_{m}} , \bar{M}_{\dot{\beta} \dot{\gamma}}] &= 2 \, m! \, i \epsilon_{\dot{\alpha}_{1} (\dot{\beta}} T_{|\alpha_{1}\ldots \alpha_{n}| \dot{\gamma}) \dot{\alpha}_{2} \ldots \dot{\alpha}_{m}} \,,
\end{align}
where we have explicitly symmetrised in $(\beta, \gamma)$ or $(\dot{\beta},\dot{\gamma})$ with weight one, where necessary. In these and all following equations, the symmetrisation with weight one of groups of indices such as $\alpha_{1}, \ldots, \alpha_{n}$ will be implicit (and similarly for the dotted indices).


\subsection{Single scalar Goldstone} \label{realscalar}

We begin with a single scalar Goldstone where all non-linear generators are fully symmetric and traceless, as argued in section \ref{Spacetime}. We denote the $n^{\text{th}}$ order generator in the inverse Higgs tree by $G_{n} \equiv G_{\alpha_{1},\ldots,\alpha_{n}\dot{\alpha}_{1},\ldots \dot{\alpha}_{n}}$ where $n = 0, 1, \ldots Z$, i.e. we include generators up to a finite order $Z$ with $G_{0}$ the essential. These generators are fully symmetric in the sets $(\alpha_{1},\ldots, \alpha_{n})$, $(\dot{\alpha}_{1},\ldots, \dot{\alpha}_{n})$ since they correspond to symmetric traceless Lorentz tensors. 

The appearance of non-linear generators in $[P_{\gamma \dot{\gamma}}, G_{\alpha_1 \ldots \alpha_n \dot{\alpha}_{1} \ldots \dot{\alpha}_{n}}] \equiv [P_{\gamma \dot{\gamma}},G_{n}]$ is fixed by our above analysis of inverse Higgs trees while the commutator between two non-linear generators remains unconstrained. We have
\begin{align} \label{realscalartree}
 [P_{\gamma \dot{\gamma}},G_{n}] &= \tfrac{1}{2} i \epsilon_{\gamma \alpha_1} \epsilon_{\dot{\gamma}\dot{\alpha}_1} G_{\alpha_2 \ldots \alpha_n \dot{\alpha}_2 \ldots \dot{\alpha}_n} \nonumber \\ &+ i A P_{\gamma \dot{\gamma}}  \quad (\text{only for}~n = 0) \nonumber  \\
&+B \epsilon_{\gamma \alpha_1}\bar{M}_{\dot{\gamma} \dot{\alpha}_1} - \bar{B}\epsilon_{\dot{\gamma} \dot{\alpha}_1} M_{\gamma \alpha_1} \,, \quad (\text{only for}~n = 1)
\end{align}
where $A$ and $B$ are respectively real and complex parameters. The fact that $B$ is complex suggests that there are two different Lorentz structures involving the Lorentz generators. In $SO(1,3)$ notation this is clearly the case, since we can write down both $M_{\mu\nu}$ and $\epsilon_{\mu\nu\rho\sigma}M^{\rho\sigma}$ on the right-hand side when $n=1$.
We could have also added a term of the form $\epsilon_{\gamma \alpha_{1}} \epsilon_{\dot{\gamma}\dot{\alpha_{1}}}P_{\alpha_{2}\dot{\alpha_{2}}}$ in the $[P_{\gamma \dot{\gamma}},G_{2}]$ commutator, but this can always be removed by a change of basis. The general form of the $ [G_{\alpha_1 \ldots \alpha_m \dot{\alpha}_{1} \ldots \dot{\alpha}_{m}}, G_{\beta_1 \ldots \beta_{n} \dot{\beta}_{1} \ldots \dot{\beta}_{n}}] \equiv [G_{m}, G_{n}]$ commutators is:
\begin{align}
[G_{m}, G_{n}]  &= \sum_{k = 0}^{n} i C_{k}^{(m,n)} \prod^{k}_{q=1} \epsilon_{\alpha_q \beta_q} \epsilon_{\dot{\alpha}_q \dot{\beta}_q} G_{\alpha_{k +1} \ldots \alpha_m \beta_{k + 1} \ldots \beta_n \dot{\alpha}_{k +1} \ldots \dot{\alpha}_{m} \dot{\beta}_{k + 1} \ldots \dot{\beta}_{n}}  \nonumber \\ 
&+ i D^m  \prod^{m-1}_{q=1} \epsilon_{\alpha_q \beta_q} \epsilon_{\dot{\alpha}_q \dot{\beta}_q} P_{\alpha_m \dot{\alpha}_m}  \quad (\text{only for}~m = n + 1) \nonumber \\
&+  \prod^{m-1}_{q=1} \epsilon_{\alpha_q \beta_q} \epsilon_{\dot{\alpha}_q \dot{\beta}_q}(E^{m} \epsilon_{\alpha_m \beta_m} \bar{M}_{\dot{\alpha}_m \dot{\beta}_m} - \bar{E}^m \epsilon_{\dot{\alpha}_m \dot{\beta}_m} M_{\alpha_m \beta_m}) \,, \quad (\text{only for}~m = n) 
\end{align}
where $C_{k}^{(m,n)}$ and $D^{m}$ are real parameters and $E^{m}$ are complex parameters. Note that $C_{k}^{(m,n)} = 0$ if $2k < (n+m-Z)$.  We have also assumed in the above that $m \geq n$ without loss of generality. 

We now constrain the form of these commutators using the remaining Jacobi identities. We consider the two cases of $Z \leq 2$ and $Z \geq 3$ separately, since the former has already been computed in \cite{LieAlgebraicScalar} (and was confirmed by our own analysis of this case).

\subsubsection*{$\bullet ~ Z \leq 2$} 
Up to and including two inverse Higgs relations, there are two branches of solutions depending on whether $A$ vanishes or not. This distinguishes between the cases where the essential generator corresponds to a shift or a scaling symmetry. 

For $A=0$ we also have $B=0$ and $G_{0}$ generates a shift symmetry for the scalar. There are two distinct algebras up to first-order ($Z=1$), with one arising as a singular contraction of the other. These correspond to the five-dimensional Poincar\'{e} algebra and its Galilean contraction. They are respectively non-linearly realised by the scalar DBI action \cite{BI,Dirac} and the Galileons \cite{Galileon}. At the level of transformation rules, the scalar DBI  transformation rule has field-dependence while this is lost in the Galilean contraction, where the transformation rule is reduced to a first order extended shift symmetry. As discussed, this field-dependence is responsible for the scalar DBI being an exceptional EFT \cite{SoftLimits2}. Both EFTs have a quadratic scaling in soft scattering amplitudes, which is related to the fact that in each case we only need to impose a single inverse Higgs constraint (to remove the inessential vector). We refer the reader to \cite{LieAlgebraicScalar} for full details on the transformation rules and algebras but let us state here that schematically the five-dimensional Poincar\'{e} algebra has 
\begin{align} 
[G_{0},G_{1}] = a P, \qquad [G_{1},G_{1}] = a M \,, \label{PM-RHS}
\end{align}
which will be a recurring theme in what follows\footnote{We note that this also includes scalar anti-DBI where the non-linearly realised algebra has two time-like directions. Whether the exceptional EFT is scalar DBI or scalar anti-DBI depends on the sign of the $a$.}.

In the presence of $G_{2}$, this set of generators again leads to two distinct algebras with one a contraction of the other. Both require the sub-algebra up to order $Z=1$ to be that of the contracted five-dimensional Poincar\'{e}. The uncontracted $Z=2$ algebra is that of the Special Galileon \cite{HinterbichlerExtended, GeometryGal} which has non-vanishing commutators between non-linear generators. The contraction again loses this property, thereby reducing the $G_{2}$ transformation rule to a second order extended shift symmetry. The Special Galileon is an exceptional EFT due to the field-dependence in the transformation rule generated by $G_{2}$. Since in both cases we need to impose two inverse Higgs constraints, both algebras lead to EFTs with a cubic soft scaling in scattering amplitudes. Again, we refer the reader to \cite{LieAlgebraicScalar,HinterbichlerExtended, GeometryGal} for full details on the Special Galileon algebra. However, let us state that the non-zero commutators between non-linear generators are of the form\footnote{Again the parameter $b$ can be positive or negative, similar to (anti-)DBI.}
\begin{align}
[G_{1},G_{2}] = b P, \qquad [G_{2},G_{2}] = b M.
\end{align}
Note the close similarity between the structure of these commutators and those in \eqref{PM-RHS}.

For $A \neq 0$, $G_{0}$ is the generator of dilatations. Jacobi identities ensure that the algebra up to first-order is that of the four-dimensional conformal algebra. It is not possible to extend the conformal group with the addition of $G_{2}$ apart from in two space-time dimensions (see for \cite{SymmetricSuperfluids} more details.). Due to the lack of shift symmetry and Adler's zero for the scalar, there is no sense in which the resulting EFT of the dilaton has an enhanced soft limit. We note that there are two well known bases for the conformal algebra but both give rise to identical EFTs\footnote{This is not always guaranteed in the presence of inverse Higgs constraints \cite{KRS}.}  \cite{Mapping}. 

\subsubsection*{$\bullet ~Z \geq 3$}

We now turn our attention to the case with more than two inverse Higgs relations, which is yet to be studied. We begin with the Jacobi identity which involves two copies of translations. Similar to the case with two inverse Higgs relations, this immediately fixes $A=B = 0$ leaving only $[P,G_{n}] = G_{n-1}$ as required to satisfy inverse Higgs relations. Next we consider the Jacobi identity with one copy of translations and two non-linear generators. By projecting onto different Lorentz structures, we find that all other parameters are also forced to vanish other than $D^{Z}$ and $Re(E^{Z})$, which are fixed to be proportional. We are therefore left with a single free parameter. The only non-vanishing commutators are those required by inverse Higgs constraints, and the following, schematically: 
\begin{align}\label{eq:dbistructure}
[G_{Z-1},G_{Z}]= D^{Z} P, \qquad [G_{Z},G_{Z}] = D^{Z} M \, .
\end{align}
Note that this structure is identical to the $Z=1$ and $Z=2$ cases above.

Finally, we consider the remaining Jacobi identities which involve three non-linear generators. Right away the Jacobi identity involving the generators $(G_{Z-2},G_{Z-1},G_{Z})$  fixes $D^{Z} = 0$ and therefore for $Z \geq 3$ all commutators between non-linear generators vanish. It follows that all symmetries reduce to extended shift symmetries and no further exceptional EFTs.
\vspace{0.1cm}

We have therefore proven, using only Lorentz invariance, the existence of inverse Higgs constraints, and Jacobi identities, that the only exceptional scalar EFTs are scalar DBI and the Special Galileon: exceptional theories with $\sigma >3$ do not exist. We refer the reader to \cite{SoftLimits2} for similar results derived using on-shell methods.


\subsection{Multiple scalar Goldstones} \label{multiscalar}
We now consider the case where there are $N>1$ essential scalar Goldstones. Most of our discussion on the single scalar carries over to this case. In particular, the inverse Higgs trees attached to the different scalar essentials decouple and each attains the same structure as in the previous section. We label our generators $G^i_{n} \equiv G^{i}_{\alpha_{1},\ldots,\alpha_{n}\dot{\alpha}_{1},\ldots \dot{\alpha}_{n}}$ according to the tree $i$ they belong to with $i=1,\ldots, N$ and their rank $n$ within that tree. Now translations act as
\begin{align}\label{eq:PGAnsatz}
[P_{\gamma \dot{\gamma}},G^{i}_{n}] &= \tfrac{1}{2} i \epsilon_{\gamma \alpha_1} \epsilon_{\dot{\gamma}\dot{\alpha}_1} G^{i}_{\alpha_2 \ldots \alpha_n \dot{\alpha}_2 \ldots \dot{\alpha}_n}  \nonumber \\ &+ i  A^i P_{\gamma \dot{\gamma}}   \quad (\text{only for}~n = 0) \nonumber  \\
&+B^i \epsilon_{\gamma \alpha_1}\bar{M}_{\dot{\gamma} \dot{\alpha}_1} - \bar{B}^i\epsilon_{\dot{\gamma} \dot{\alpha}_1} M_{\gamma \alpha_1}  \quad (\text{only for}~n = 1)\,,
\end{align}
with each tree containing a finite number $Z_i$ of inessential generators. The coefficient $A^i$ may, without loss of generality, be set to zero for all but a single non-linear scalar generator i.e. there can only be a single dilaton. The commutators $[G^i_n, G^j_m]$ are also very similar to the previous section, but coefficients now carry the appropriate extra indices
\begin{align}\label{eq:GGAnsatz}
[G^i_{m}, G^j_{n}]  &= \sum_{k=1}^{N}\sum_{w = 0}^{n} \prod^{w}_{q=1} \epsilon_{\alpha_q \beta_q} \epsilon_{\dot{\alpha}_q \dot{\beta}_q} i  D_{w}^{(i,m;j,n)k} G^{k}_{\alpha_{w +1} \ldots \alpha_m \beta_{w + 1} \ldots \beta_n \dot{\alpha}_{w +1} \ldots \dot{\alpha}_{m} \dot{\beta}_{w + 1} \ldots \dot{\beta}_{n}}  \nonumber \\ 
&+ i F^{ij,m}  \prod^{m-1}_{q=1} \epsilon_{\alpha_q \beta_q} \epsilon_{\dot{\alpha}_q \dot{\beta}_q} P_{\alpha_m \dot{\alpha}_m}  \quad (\text{only for}~m = n + 1) \nonumber \\
&+  \prod^{m-1}_{q=1} \epsilon_{\alpha_q \beta_q} \epsilon_{\dot{\alpha}_q \dot{\beta}_q}(H^{ij,m} \epsilon_{\alpha_m \beta_m} \bar{M}_{\dot{\alpha}_m \dot{\beta}_m} - \bar{H}^{ij,m} \epsilon_{\dot{\alpha}_m \dot{\beta}_m} M_{\alpha_m \beta_m})  \quad (\text{only for}~m = n) \nonumber \\
&+ \prod^{m}_{q=1} \epsilon_{\alpha_q \beta_q} \epsilon_{\dot{\alpha}_q \dot{\beta}_q} i X^{ij,m} \quad (\text{only for}~m = n) \, ,
\end{align}
where we have taken $m \geq n$. The parameters $D_{w}^{(i,m;j,n)k}$ and $F^{ij,m}$ are real, whereas $H^{ij,m}$ are complex in general. The linear scalar generators $X^{ij,m}$ are defined by this commutation relation. Since they are linearly realised, they commute with translations, form a sub-algebra, and their commutation relations with non-linear generators can only produce non-linear generators.

When $m = n$, the right-hand side needs to be anti-symmetric under the simultaneous exchange of the Lorentz indices on $G^i_m$ and $G^j_m$ and the tree labels $i$ and $j$. This imposes the conditions:
\begin{equation}
D_{w}^{(i,n;j,n)k} = -D_{w}^{(j,n;i,n)k} \,, \quad H^{ij,m} = H^{ji,m} \,, \quad X^{ij,m} = -X^{ji,m} \,.
\end{equation}
In particular, when there are two scalar essentials ($N=2$), there is only a single linear scalar generator at each order: $X^{ij,m} \equiv X^{m}$. We also have $D_{w}^{(i,m;j,n)k} = 0$ when $2 w < (n+m-Z_k)$.

We now consider the following cases separately: firstly we investigate the case where there are no inverse Higgs constraints i.e. $Z_{i} = 0$ for each tree. We then consider the case where no tree involves more than a single inessential Goldstone i.e $Z_{max} = 1$. Finally, we consider the case where at least one essential Goldstone contains at least two inessentials in its inverse Higgs tree i.e. $Z_{max} \geq 2$. 

\subsubsection*{$\bullet ~Z_{max}=0$}
In the case $Z_{max} = 0$, where no tree has any inessentials, all generators other than translations and Lorentz transformations are scalars. We collectively denote them as $(Y^i,D)$ for simplicity, where $D$ is the generator of dilatations. We assume that $D$ is a non-linear generator while $Y^{i}$ includes both linear and non-linear generators. After imposing all the constraints from Jacobi identities we have
\begin{equation}
[P_{\alpha \dot{\alpha}}, Y^i] = 0, \quad [P_{\alpha \dot{\alpha}}, D] = i P_{\alpha \dot{\alpha}} \,,
\end{equation}
and
\begin{align}
[Y^i, Y^j] = i D^{ij}{}_{k} Y^k, \quad [D, Y^i] =i E^{i}{}_{j} Y^j \, ,
\end{align}
with the constraints
\begin{align} \label{eq:dilatonconstraint}
D^{[i j}{}_{k} E^{k]}{}_{l} = 0, \qquad  D^{[i j}{}_{l} D^{k] l}{}_{m} = 0 \,. 
\end{align}
In the presence of dilatations, each $Y^{i}$ can therefore have a non-trivial scalar weight.  These algebras of course include the cases of a number of shift symmetric scalars coupled to a dilaton.

In general the scalars of these theories are said to span a non-linear sigma-model. In the two-scalar cases they also include the well known coset spaces
\begin{align}
\frac{SO(3)}{SO(2)}, \qquad \frac{SO(1,2)}{SO(2)} \,,
\end{align} 
which appear often in the inflationary literature\footnote{They also include the algebra of the scaling superfluid presented in \cite{SymmetricSuperfluids} and we refer the reader there for more details.}, e.g.~as  $\alpha$-attractors \cite{AlphaAttractors}. Given our definition of exceptional EFTs, such non-linear sigma-models define an exceptional EFT since the two-derivative action, which includes interactions, is completely fixed by symmetry. Indeed, the transformation rules include field-dependent pieces. 

\subsubsection*{$\bullet ~Z_{max}=1$}
We now turn to the case where $Z_{max} = 1$, where each tree has at most one inessential generator. Here we find it useful to separate the calculation into two sub-cases: in the first we do not allow any inessential generators in the dilaton's inverse Higgs tree (if the dilaton exists in the first place), while in the second case we do allow for that vector inessential. The Jacobi identity $(P, P, G)$ imposes $[P, K] \propto D + M$ for the vector inessential $K$ of the dilaton $D$. This means that $K$ necessarily generates a type of special conformal transformation. Because the case $Z_{max} = 1$ without the dilaton was considered in \cite{Brauner2,LieAlgebraicScalar}, we will focus on what changes when the dilaton is included.

We define our generators as follows: the scalar generator $D$ is the dilaton. The generators $G^i$ are scalars which have inessential vectors $G^i_{\alpha \dot{\alpha}}$. Furthermore, we have scalars $X^I$ which do not fit into the previous two categories i.e. they can be linearly realised or correspond to scalar Goldstones with empty inverse Higgs trees.

\subsubsection*{Without special conformal transformations}
In the first sub-case, after we have imposed all the constraints from Jacobi identities on the Ansatz \eqref{eq:PGAnsatz}\eqref{eq:GGAnsatz}, the part of the algebra that does not involve the dilatations reduces to
\begin{align}\label{eq:Zmaxone}
&[P_{\gamma \dot{\gamma}}, G^i_{\alpha \dot{\alpha}} ] = \tfrac{1}{2} i \epsilon_{\gamma \alpha} \epsilon_{\dot{\gamma} \dot{\alpha}} G^i \,, \quad [G^i_{\alpha \dot{\alpha}}, G^j] = i  H^{ij} P_{\alpha \dot{\alpha}} \nonumber \\
&[G^i_{\alpha \dot{\alpha}}, G^j_{\beta \dot{\beta}}] = 4i H^{ij}(\epsilon_{\alpha \beta} \bar{M}_{\dot{\alpha} \dot{\beta}} + \epsilon_{\dot{\alpha}\dot{\beta}} M_{\alpha \beta}) + i \epsilon_{\alpha \beta} \epsilon_{\dot{\alpha}\dot{\beta}}Y^{ij} \,, \quad [X^I, X^J] = i f^{I J}{}_{K} X^K \,, \nonumber \\
&[X^I, G^i] = iB^{iI}{}_{j} G^{j} + iD^{iI}{}_{J} X^J, \quad  [X^I, G^i_{\alpha \dot{\alpha}}] =  i B^{iI}{}_{j} G^j_{\alpha \dot{\alpha}} + iC^{iI} P_{\alpha \dot{\alpha}}\, .
\end{align}
We arrive at this result by eliminating the generator $D$ from the right-hand side of each of the above commutators. Then the calculation reduces to the case considered in \cite{Brauner2}. We will consider the commutators involving $D$ in a moment. 

The commutator of the two inessential vectors defines the scalars $Y^{ij}$. These generators are not independent from $G^i$ and $X^I$. In general, a linear combination of $G^i$ and $X^I$ generators can appear on the right-hand side of the commutator\footnote{The commutation relations of the $Y^{ij}$ generators are fixed by Jacobi identities, but that does not identify the linear combination of $G^i$ and $X^I$ generators that appears in the commutator. In particular, we can always add a central charge $C$. When linearly realised, this does not change the transformation laws or invariants derived from the coset construction. When non-linearly realised, such an extension makes it impossible to impose inverse Higgs relations in all examples we know of, but we do not know of a proof that this always happens.}. There are several additional constraints on the coefficients in \eqref{eq:Zmaxone}. The full list appears in \cite{LieAlgebraicScalar}\footnote{Note that the coefficients $a_{A}^i$ and $e_i^A$ of equation (4.1) in \cite{LieAlgebraicScalar} are fixed (up to a basis change) by the inverse Higgs trees to be diagonal and zero, respectively. We have furthermore divided the scalar sectors in a different way, which is why we are able to remove the term $[G, G_{\alpha \dot{\alpha}}] \propto G_{\alpha \dot{\alpha}}$.} and we refer the reader there for full details. Here we simply comment on the structure of the solutions. The matrix $H^{ij}$ can be made diagonal by a basis change. Then, the non-zero elements can be made $1$ or $-1$ by rescaling the $G^i$ generators. The positive or negative choices determine the signature of the metric in the higher-dimensional spacetime. Vanishing diagonal elements correspond to Galileons which can be coupled to the DBI scalars. Equivalently, they are the result of an In\"onu-Wigner contraction of a subset of the higher-dimensional coordinates. The matrix $H^{ij}$ fixes the commutation relations of the $Y^{ij}$ with themselves and with $G^i_{\alpha \dot{\alpha}}$. In the case that $H^{ij} = \delta^{ij}$, the symmetry algebra contains a factor $ISO(1,3+N)$ (where $N$ is the number of inverse Higgs trees) and $Y^{ij}$ generate the $SO(N)$ subgroup. This algebra leads to multi-DBI \cite{multiDBI}.

We now consider the commutations relations between dilatations $D$ and the other generators. We fix the dilatation weight of the translation generator to unity: $[P_{\gamma \dot{\gamma}}, D] = i P_{\gamma \dot{\gamma}}$. The remaining commutation relations are
\begin{align}
&[D, G^i] = iB^{i}{}_j G^j, \quad [D, G_{\alpha \dot{\alpha}}^i] = i(B^{i}_{j} - \delta^{i}_{j})G_{\alpha \dot{\alpha}}^j + i J^i P_{\alpha \dot{\alpha}} \,, \nonumber \\
&[D, X^I] = i S^{I}{}_i G^i + i T^{I}{}_J X^J \,,
\end{align}
with constraints
\begin{align}
&T^{[ij] J} = 0, \quad S^{[ij]}{}_k = J^{[i}\delta^{j]}{}_k, \quad H^{ij} = -(B^{ik} - \delta^{ik}) H^{k j} \, .
\end{align}
The pair of indices $ij$ on the $T$ and $S$ coefficients is a special case of the general index $I$, as before. In addition to these, we have the usual constraint of the form \eqref{eq:dilatonconstraint} that relates the structure constants of the scalar subalgebra to the dilatation weights. Finally, the last constraint fixes the weights of the higher-dimensional translation and boost generators. Taking $H^{ij}$ diagonal, it follows that the DBI scalars have zero scaling weight. The weights of the Galileon directions are not fixed by this equation.

We note that, similar to the $Z_{max} = 0$ case, there are many different solutions to the Jacobi identities for different choices of the generator content. However, the structure of those solutions is very simple: in every case the vectors generate the symmetry algebra of a higher-dimensional space, coupled to a set of Galileons. Furthermore, one can add a dilaton and some internal coset space $G / H$. The dilatation weights of the DBI scalars are fixed and their representation under the internal coset space must satisfy the constraints of \cite{LieAlgebraicScalar}. 

Let us explain in more detail why we can couple a DBI scalar to a Galileon. Consider the six-dimensional Poincar\'{e} algebra with generator content: $P_{\mu},M_{\mu\nu},P_{5},P_{6},M_{\mu 5},M_{\mu 6},M_{56}$ where $(5,6)$ refer to the two extra dimensions. In the usual construction of multi-DBI, $P_{5},P_{6}$ correspond to the two essential scalars, $M_{\mu 5},M_{\mu 6}$ are the two inessential vectors appearing in each inverse Higgs tree and $M_{56}$ is a linearly realised $SO(2)$ between the two scalars. However, we can take a singular contraction by rescaling $P_{6} \rightarrow \omega P_{6},M_{\mu 6} \rightarrow \omega M_{\mu 6}$ and $M_{56} \rightarrow \omega M_{56}$ with $\omega \rightarrow \infty$ such that the scalar corresponding to $P_{5}$ is a DBI scalar with schematically $[M_{\mu 5},M_{\nu 5}] = M_{\mu\nu}$, $[P_{5},M_{\mu 5}] = P_{\mu}$ while the scalar corresponding to $P_{6}$ reduces to a Galileon with $[M_{\mu 6},M_{\nu 6}] = 0$, $[P_{6},M_{\mu 6}] =0$. We therefore have a DBI scalar coupled to a Galileon but let us stress that the presence of $M_{56}$ is crucial since $[M_{\mu 5},M_{\nu 6}] = \eta_{\mu\nu}M_{56}$, both before and after the contraction.
\subsubsection*{With special conformal transformations}
Next we consider the case where $D$ has a vector inessential $K_{\alpha \dot{\alpha}}$. We immediately have
\begin{equation}
[P_{\gamma \dot{\gamma}}, K_{\alpha \dot{\alpha}}] = -i \epsilon_{\gamma \alpha} \epsilon_{\dot{\gamma} \dot{\alpha}} D + \tfrac{i}{2} \epsilon_{\gamma \alpha} \bar{M}_{\dot{\gamma}\dot{\alpha}} + \tfrac{i}{2} \epsilon_{\dot{\gamma}\dot{\alpha}} M_{\gamma \alpha} \, .
\end{equation}
The Jacobi identity $(P, K, D)$ fixes the subalgebra spanned by $D$, $K_{\alpha \dot{\alpha}}$ and the Poincar\'e generators to the ordinary $\text{AdS}_5$ algebra. We have checked that it is not possible to extend this algebra with other inessential vector generators. This is an unsurprising result, because the $\text{AdS}_{4 + n}$ algebra satisfies $Z_{max} = 2$ for $n > 1$. It is not possible to truncate these algebras to a $Z_{max} = 1$ component. We will return to the higher-dimensional Anti-de Sitter algebras in the following subsection.

\subsubsection*{$\bullet ~Z_{max} \geq 2$}
We now consider the case where at least one of the trees, say $i = 1$, includes a second-order inessential generator i.e. $Z_{1} \geq 2$. We do not assume anything about the other trees. However, their structure will be constrained by Jacobi identities. We begin with the Jacobi identities involving two copies of translations and one non-linear generator since these Jacobi identities do not mix the different trees. It is simple to see that each tree which includes a second-order generator must have $A=B=0$ i.e. an essential scalar generator can only generate dilatations if its tree has at most one non-linear generator (which of course corresponds to special conformal transformations).

We now move on to Jacobi identities involving one translation generator and two non-linear generators from any tree i.e. $(P, G_{m}^i, G_{n}^j)$. This tells us that $B^{i} = 0$ for all trees, so any algebras involving at least one second-order, traceless generator cannot form an extension the conformal algebra. This means that if the dilaton exists as an essential Goldstone, it cannot have any inessentials in its own inverse Higgs tree. For this reason we will assume that the dilaton is not included for the moment and come back to that possibility later. The remaining constraints tells us the following:

\begin{itemize}
\item $H^{ij,Z_i} = 4i F^{ij,Z_j}$ if $i$ and $j$ label two trees with $Z_i = Z_j$. All other $H$ and $F$ coefficients are zero. This structure strongly resembles \eqref{eq:dbistructure}.
\item The linear scalars $X^{ij,m}$ can only appear if $Z_i = Z_j = m$.
\item The appearance of non-linear generators $G^k$ in the commutators $[G^i, G^j]$ is also highly constrained. The only allowed structure is $[G^i_{Z^i}, G^j_{m}] \supset G^k_{m - Z^i}$ where $m \geq Z^i$.
\end{itemize}

We now consider Jacobi identities with three non-linear generators, $(G^i, G^j, G^k)$. We begin by taking the second-order generator $G^1_{\alpha_1 \alpha_2 \dot{\alpha}_1 \dot{\alpha}_2}$ from the $i = 1$ tree, together with two vector inessentials $G^{j}_{\beta \dot{\beta}}, \, G^{k}_{\gamma \dot{\gamma}}$ from any trees in the algebra. From inspecting the terms proportional to $G^1_{\alpha_1 \alpha_2 \dot{\alpha}_1 \dot{\alpha}_2}$ we obtain $H^{jk,1} = 0$, telling us that any tree with $Z=1$ cannot realise the DBI structure \eqref{eq:dbistructure}. We also see that any scalar generator which has a non-zero commutator with $G^1_{\alpha_1 \alpha_2 \dot{\alpha}_1 \dot{\alpha}_2}$ cannot appear in any commutator involving two vectors.

From this Jacobi identity we can also infer that it is impossible to couple several Special Galileons. Indeed, if we take $i=j=1$ and $Z_k = 2$, there are two terms proportional to non-linear scalars given by
\begin{equation}
iF^{ik, 2}\epsilon_{\alpha_1 \gamma} \epsilon_{\dot{\alpha}_1 \dot{\gamma}} \epsilon_{\alpha_2 \beta} \epsilon_{\dot{\alpha}_2 \dot{\beta}} G_0^j - i F^{ij, 2} \epsilon_{\alpha_1 \beta_1} \epsilon_{\dot{\alpha}_1 \dot{\beta}} \epsilon_{\alpha_2 \gamma} \epsilon_{\dot{\alpha}_2 \dot{\gamma}} G_0^k \,
\end{equation}
with symmetrisation over the $\alpha$ indices assumed, as usual. These terms only cancel when $i = j = k$. Therefore, at most one Special Galileon can exist at once. The same result, for the case of two scalar fields, was found from amplitude methods in \cite{SoftBootstrap}.

The terms proportional to non-linear scalars impose important constraints as well. Taking the trees $i$ and $k$ to be the same, $i = k = 1$, and $j$ to be some tree that obeys $Z_j = 1$, we find
\begin{equation}\label{eq:specGal}
F^{11, 2} = - 2 D^ {(1 2; j 1) m} D^{(1 1; m 1) j} \, .
\end{equation}
The coefficient on the left-hand side $F^{11, 2}$ determines whether the Special Galileon structure \eqref{eq:dbistructure} is realised by the tree $i = 1$. The coefficients on the right-hand side tell us whether the commutator $[G^1_2, G^j_1]$ (where the subscript refers to the number of Lorentz indices) contains a vector $V$ which satisfies $[G^1_1, V] \propto G_0^j$. We will now shown that no such vector $V$ can exist. To do so we inspect the Jacobi identity involving three vector inessentials, two of them from trees with $Z \geq 2$ and one of them from a tree with $Z = 1$. Inserting the constraint $[G^i_{Z^i}, G^j_{m}] \supset G^k_{m - Z^i}$, we find
\begin{equation}
[G^i_1, G^j_1] \not \supset G_0^k \quad (\text{if } Z_i = 1, \, Z_j = 2) \, ,
\end{equation}
which implies that the right-hand side of \eqref{eq:specGal} is equal to zero. Therefore, the existence of a Special Galileon kills the possibility of other inessential Goldstones. They can only be coupled to scalar Goldstones that do not realise additional inessential symmetries.

We have therefore seen that if the EFT includes a Special Galileon, it must be the only one and can only couple to other scalars which have empty inverse Higgs trees. This is in stark contrast to the $Z_{max}=1$ case where we can have multi-DBI. Furthermore, if the EFT contains any scalar which has $Z \geq 3$, only extended shift symmetries are possible for each scalar. In that case the different $Z_{i}$ are unconstrained, i.e. each tree can have top rank generators of different Lorentz representation. In conclusion, the $Z_{max} \geq 2$ coset spaces are
\begin{equation}
 \left( \frac{SG(1,3 + 1)}{SO(1,3)} \times \frac{G'}{H'} \right) \,, \quad \left( \frac{\text{Extended shifts}}{SO(1,3)} \rtimes \frac{G'}{H'} \right)  \, ,
\end{equation}
where $SG$ refers to the special Galileon algebra and $G'/H$ are internal coset spaces\footnote{Note that as is usual in the coset construction for spacetime symmetries, the translation generators are included in $G/H$ rather than $H$. So in these coset spaces the linearly realised subalgebra is therefore $\text{Lorentz} \times H'$.}. Note that in the latter case, the extended shift algebra can form a non-trivial representation of $G'$. As there is a single scalar, this is impossible for the special Galileon algebra.

\subsubsection*{Off-diagonal generators and dilatons} 
An important caveat concerns our restriction to purely symmetric and traceless representations in inverse Higgs trees, i.e.~from $Z_{max}=2$ onward. As explained in chapter \ref{Spacetime}, these are the unique transformations that leave the kinetic terms invariant provided the latter are the lowest order terms in a derivative expansion. This is a general statement in the absence of a dilatation Goldstone. In the presence of a dilaton Goldstone, however, non-linearly realised symmetries may relate the dilaton potential to kinetic terms.

An interesting example of this possibility is provided by the AdS$_{4+n}$ algebra, that can be written as
\begin{equation*}
\begin{aligned}[c]
&[P_A,D] = P_A \,, \\
&[{K}_A,D] = - {K}_A  + P_A \,,\\
&[P_A,{K}_B] = 2 M_{AB}  + 2\eta_{AB}D\,, \\
&[{K}_A,{K}_B] = 2M_{AB} \,,
\end{aligned}
\qquad
\begin{aligned}[c]
&[M_{AB},P_C] = \eta_{AC}P_B - \eta_{BC}P_A \,, \\
&[M_{AB},{K}_C] = \eta_{AC}{K}_B - \eta_{BC}{K}_A \,,  \\
&[M_{AB},M_{CD}] = \eta_{AC}M_{BD} - \eta_{BC}M_{AD} + \eta_{BD}M_{AC} - \eta_{AD}M_{BC} \,,
\end{aligned}
\end{equation*}
with $A = (\mu, i_2, \ldots i_n)$. For $n=1$, this only involves a dilaton Goldstone and its inverse Higgs vector of special conformal transformations, as discussed in section \ref{realscalar}. However, when $n \geq 2$ this set-up is augmented with $n-1$ trees consisting of an axion Goldstone, its inverse Higgs vector of Lorentz boosts, as well as a $Z=2$ scalar arising from special conformal transformations in the higher-dimensional directions\footnote{This algebra allows for (at least) two inequivalent In\"onu-Wigner contractions, leading either to the Poincar\'e or the Galileon algebra, as discussed in the single-field case in \cite{Reunited}. Importantly, the contraction that gives rise to Poincar\'e does not preserve the structure of the inverse Higgs relations.}. It is discussed in \cite{KRS2} how the lowest order invariant that includes the kinetic terms also generates a potential term for the dilaton. This combination allows for the $Z=2$ scalar in the axion trees, which was ruled out in the general discussion above under the assumption of having shift symmetries and no dilatations. 

Crucially, the combination of special conformal transformations in one inverse Higgs tree, and a $Z=2$ symmetric traceless generator in another, was ruled out in the above. Moreover, the inclusion of any off-diagonal generators other than the $Z=2$ scalar requires the $Z=2$ symmetric traceless one, as discussed in section \ref{Spacetime}. This implies that the above exception based on conformal symmetry is the unique one; adding additional Goldstone modes to this can only give rise to higher-dimensional Anti-de Sitter algebras. However,  these would not be exceptional EFTs with soft limits as defined in this paper, due to the different implications of scaling symmetries.

Finally, let us mention that we can still include the dilaton with an empty inverse Higgs tree, i.e. without special conformal transformations. Here the algebras are the ones discussed above but each generator can have a non-vanishing scaling weight with generalisations of the \eqref{eq:dilatonconstraint} constraints. 

\subsection{Spin-$1/2$ Goldstones} \label{fermion}

We now study the case where the essential Goldstones are $N$ spin-1/2 fermions $\chi_{\alpha}^i$, with $i = 1, \ldots, N$. Any higher-order generators we add to this algebra to realise more symmetries on the essential fermions must also be fermionic, as they are related to the essential fermionic generators by space-time translations. Moreover, since the anti-commutator between two fermionic generators can only give rise to bosonic generators (in this case only linear generators), the algebras at every order will always form subalgebras. Note that this is very different to the bosonic case where there is much more freedom in a commutator between two non-linear generators, as illustrated by the discussion above.

Given our discussion in \ref{stepI} we know that the inverse Higgs tree for each fermion decouples. Section \ref{stepII} tells us that if the essential fermions are to have canonical propagators, we can only add a single non-linear generator at order $n$ in each inverse Higgs tree which has spin-$(n+1/2)$. We again consider non-linear generators up to finite order $Z_{i}$, allowing for different top levels for each fermion, denoted $\chi^{i}_{n} \equiv \chi^i_{\alpha_1 \ldots \alpha_{n+1} \dot{\alpha}_1 \ldots \dot{\alpha}_{n}}$ with Hermitian conjugate  $\bar{\chi}^{i}_{n} \equiv \bar{\chi}^i_{\alpha_{1}\ldots \alpha_{n} \dot{\alpha}_1 \ldots \dot{\alpha}_{n+1}}$, where $n = 0, \ldots Z_{i}$. 
The inverse Higgs tree fixes the commutators between translations and non-linear generators to be 
\begin{align}
&[P_{\gamma \dot{\gamma}}, \chi^i_{n}] = i \epsilon_{\gamma \alpha_1} \epsilon_{\dot{\gamma} \dot{\alpha}_1} \chi^i_{\alpha_2 \ldots \alpha_{n+1} \dot{\alpha}_2 \ldots \dot{\alpha}_{n}} \, ,
\end{align}
while commutators between two non-linear generators are only constrained by the linear symmetries. We have 
\begin{align}
\{\chi^i_{m}, \chi^j_{n} \} &= A^{(ij,m)}  \prod^{m}_{q=1} \epsilon_{\alpha_q \beta_q} \epsilon_{\dot{\alpha}_q \dot{\beta}_q} M_{\alpha_{m+1} \beta_{m+1}}  \quad  (m=n) \nonumber \\
&+B^{(ij,m)} \prod^{m-1}_{q=1} \epsilon_{\alpha_q \beta_q} \epsilon_{\dot{\alpha}_q \dot{\beta}_q}  \epsilon_{\alpha_{m} \beta_{m}}  \epsilon_{\alpha_{m+1} \beta_{m+1}} \bar{M}_{\dot{\alpha}_{m} \dot{\beta}_{m}}  \quad (m=n) \nonumber \\
&+ C^{(ij,m)}  \prod^{m-1}_{q=1} \epsilon_{\alpha_q \beta_q} \epsilon_{\dot{\alpha}_q \dot{\beta}_q} \epsilon_{\alpha_{m}\beta_{m}} P_{\alpha_{m+1} \dot{\alpha}_{m}}, \quad (m = n + 1)
\end{align}
with $m \geq n$ and complex parameters, and 
\begin{align}
\{\chi^{i}_{m}, \bar{\chi}^{j}_{n} \} &= D^{(ij,m)}  \prod^{m-1}_{q=1} \epsilon_{\alpha_q \beta_q} \epsilon_{\dot{\alpha}_q \dot{\beta}_q} \epsilon_{\dot{\alpha}_{m}\dot{\beta}_{m}}  M_{\alpha_{m} \alpha_{m+1}} \quad (m = n + 1)\\
 &+ E^{(ij,n)}  \prod^{n-1}_{q=1} \epsilon_{\alpha_q \beta_q} \epsilon_{\dot{\alpha}_q \dot{\beta}_q} \epsilon_{\alpha_{n} \beta_{n}}  \bar{M}_{\dot{\beta}_{n } \dot{\beta}_{n+1}}  \quad (n = m + 1)\\
&+F^{(ij,m)} \prod^{m}_{q=1} \epsilon_{\alpha_q \beta_q} \epsilon_{\dot{\alpha}_q \dot{\beta}_q} P_{\alpha_{m+1} \dot{\beta}_{m+1}} \, , \quad (m=n)
\end{align}
where again all parameters are complex. We now consider   $Z_{i}=0$ and $Z_{max} \geq 1$ separately.

\subsubsection*{$\bullet ~Z_{i}=0$}
First consider the case without any inessential generators, where the results are well known. The allowed algebras correspond to $N$-extended super-Poincar\'{e} and Inonu-Wigner contractions thereof. The only non-trivial and non-vanishing commutators in the uncontracted algebra are\footnote{The appearance of $\delta^{ij}$ is guaranteed by positivity in Hilbert space. This is a necessary requirement in any linear realisations of the symmetry algebra, but not in non-linear realisations as the currents don't integrate into well-defined charges in the quantum theory. Here we still assume the requirement of positivity in Hilbert space. This is a reasonable assumption if one anticipates that the non-linear realisations have a (partial) UV completion to a linearly realised theory, or to be a particular limit of such a theory.}
\begin{align} \label{SuperPoinc}
\{\chi_\alpha^{i}, \bar{\chi}^{j}_{\dot{\beta}}\} = -2 \delta^{ij}P_{\alpha \dot{\beta}} \,.
\end{align} 
At lowest order in derivatives, the EFT which non-linearly realises the $N$-extended super-Poincar\'{e} algebra is that of multi Volkov-Akulov (VA) \cite{VA}. The commutator \eqref{SuperPoinc} guarantees that the transformation rules for each fermion are field-dependent and therefore VA is an exceptional EFT with $\sigma = 1$ soft behaviour. 

In addition there are many different contractions of this algebra that give rise to new ones. We can take the limit where $\{\chi^{i}_\alpha, \bar{\chi}^{j}_{\dot{\beta}}\} = 0$ for all $i,j$ in which case all transformation rules reduce to shift symmetries for each fermion, see \cite{VA-SmallFieldLimit}. However, we do not have to perform this contraction for all $N$ generators. We can do it to none, all or any other number in between. Indeed we can realise an EFT consisting of $N_{1}$ shift symmetric fermions and $N_{2}$ VA fermions with the only constraint that $N = N_{1} + N_{2}$. The $N=1$ contracted case was studied in detail in \cite{VA-SmallFieldLimit}, where it was shown that in four-dimensions the only Wess-Zumino term one can write down is the fermion's kinetic term, i.e.~all interactions need at least one derivative per field.

\subsubsection*{$\bullet ~Z_{max} \geq 1$}
We now consider adding higher-order inessential generators with inverse Higgs relations. We allow for different top levels in each tree but we assume that at least one tree includes inessential generators. We follow exactly the same process as we did previously: we use the Jacobi identities $(P_{\mu},\chi_{m},\chi_{n})$, $(P_{\mu},\chi_{m},\bar{\chi}_{n})$ and $(\chi_{m},\chi_{m},\bar{\chi}_{n})$ and find that all free parameters must vanish. We therefore find that for $Z_{max} \geq 1$ the only non-trivial commutators are those required by inverse Higgs constraints, which results in extended shift symmetries for all the fermions. There are no other exceptional EFTs. This includes the fermionic generalisation of scalar multi-Galileons\footnote{See \cite{multigal} for a discussion on bi-Galileons.}, which are invariant under shifts linear in the coordinates. This theory, for the case of a single fermion essential, was also discovered in \cite{SoftBootstrap} using soft amplitudes. We have therefore seen that field-dependent transformation rules for the essential fermions are incompatible with inverse Higgs constraints. The only exceptional fermion EFT is that of Volkov-Akulov and its multi-field extensions.

\subsection{Including a $U(1)$ gauge vector} \label{vectoressential}

After the above full classifications for the cases of scalar or fermion Goldstone modes, we would now like to discuss a number of aspects when turning to a gauge vector instead.

The simplest possible scenario would involve only a gauge vector. A natural question is whether there are gauge theories that have non-trivial soft limits and hence non-linear symmetries. This question was answered negatively for a $U(1)$ vector in \cite{LieAlgebraicVector}. This result can be explained in a simple manner given the inverse Higgs framework of section \ref{Spacetime}. Indeed, the gauge symmetry of a vector can be seen as an infinite sequence of non-linearly realised symmetries of the form
 \begin{align}
   \delta A_\mu = u_\mu + u_{\mu \nu} x^\nu + u_{\mu \nu \rho} x^\nu x^\rho + \ldots \,, \label{gauge}
 \end{align}
where the $u_{\mu \cdots}$ parameters are symmetric and contain traces. The first non-trivial extension of this symmetry under which the field strength transforms consists of an anti-symmetric component $\delta A_\mu = b_{\mu \nu} x^\nu + \ldots$, where the dots indicate possible field-dependent terms. There are similar structures at higher powers of the coordinates that involve mixed symmetry tensors. However, these always require the two-form generator $B_{\mu\nu}$ to be included as well, and moreover the transformations up to and including linear terms always form a subalgebra \cite{LieAlgebraicVector}. It therefore suffices to investigate the implications of this finite algebra. 

In contrast to the scalar and fermion cases, it was shown in \cite{LieAlgebraicVector} that it is impossible in the vector case to have non-vanishing commutation relations between non-linear generators of this algebra\footnote{This is only true in the presence of gauge symmetries. If we drop this requirement then a single vector can, for example, non-linearly realise a double-copy of the Poincar\'{e} algebra. We expect that this theory will contain ghosts or have an infinitely strongly coupled longitudinal mode but this might require further study.}. Therefore, there are no exceptional EFTs for a single gauge vector. This was also found from an amplitude perspective in \cite{SoftBootstrap}. The only non-trivial possibility beyond gauge symmetry is therefore to have the field-independent transformation $\delta A_\mu = b_{\mu \nu} x^\nu$, which can be seen as the vector analogon of Galilean transformations. However, there are no corresponding interactions which do not introduce additional ghostly degrees of freedom \cite{Deffayet}.

The next simplest possibility would be to have a number of essential Goldstones, for instance a scalar and a vector. Given the discussion of non-linear symmetries in Section \ref{Spacetime}, the two inverse Higgs trees of these Goldstones decouple. Therefore the only unknowns are the commutators between Goldstones. We will not perform an exhaustive classification of possible algebras, but rather point out an interesting possibility. 

We start with a given exceptional algebra on the scalar side, i.e.~scalar DBI or Special Galileon and add to this the infinite sequence of gauge transformations of the vector \eqref{gauge}. Consistency of the algebra requires the vector field to transform in a specific way under the generators in the scalar's inverse Higgs tree. Consider for instance the DBI exceptional algebra with generators $X,Y_{\mu}$ spanning the scalar algebra: $X$ generates the scalar shift symmetry while the inessential vector $Y_{\mu}$ generates the Lorentz boosts in the extra dimension. Under $X$ the gauge field does not transformation while under the inessential vector $Y_{\mu}$ it has to transform as \cite{Gliozzi, Gomis}
 \begin{align}
  \delta \phi = y^\mu (x_\mu + \phi \partial_\mu \phi) \,, \qquad
 \delta A_\mu = y^\nu ( \phi \partial_\nu A_\mu + A_\nu \partial_\mu \phi ) \,,
 \end{align}
where we also included the transformation law of the scalar. This implies that the gauge vector forms a linear representation of the inessential vector of the scalar algebra and therefore specific couplings between the DBI scalar and vector are required to maintain invariance of the action. This leads to the coupled DBI theory, see \cite{Gliozzi} for more details. The special pedigree of this theory can be seen from e.g.~its higher-dimensional origin, its possible supersymmetrisation and, given the present discussion, also from the perspective of soft limits, which would be $\sigma=2$ and $\sigma=0$ for the scalar and vector, respectively. 

We now discuss a similar situation for coupling a gauge vector to the only other scalar exceptional EFT, namely, the Special Galileon. When there are no additional generators in the inverse Higgs tree of the vector (besides the infinite sequence of gauge symmetries), we have checked that the algebra on the scalar side cannot be modified by adding gauge transformations. We now have $X$ (essential scalar), $Y_{\mu}$ (inessential vector) and $Z_{\mu\nu}$ (inessential symmetric, traceless rank-2) in the scalar inverse Higgs tree and again we find that the vector only transforms under the highest of these, $Z_{\mu\nu}$. The transformations take the unique form 
 \begin{align}
  \delta \phi = z^{\mu \nu} (x_\mu x_\nu + \partial_\mu \phi \partial_\nu \phi) \,, \qquad  \delta A_\rho = 2 z^{\mu \nu} ( \partial_\mu \phi \partial_\nu A_\rho + A_\mu \partial_\nu \partial_\rho \phi) \,.
 \end{align}
Again, the gauge vector forms a linear representation of the highest inessential of the scalar algebra, in this case the symmetric traceless tensor. The same applies to the gauge transformations, which commute with $X,Y_{\mu}$ but commute with $Z_{\mu\nu}$ into a gauge transformation. We therefore have a field content consisting of a scalar that is a Special Galileon Goldstone, and a vector that transforms as a matter field. These symmetries require specific couplings between the fields in the resulting EFT which lead to soft limits of $\sigma=3$ and $\sigma =0$ for the scalar and vector, respectively. Interactions of this kind with exactly these soft limits were recently proposed in \cite{SoftBootstrap}, and therefore we expect that the above symmetry for the vector explains the non-trivial couplings found in that work but a more complete study of the possible interactions would be an interesting avenue for future work.

These two possibilities exhaust the couplings between an essential Goldstone scalar and a gauge vector, unless inessential generators other than gauge symmetries are included for the vector. We leave for future work the classification of the possible symmetries including those generators.

\section{Conclusions}

This paper provides a systematic approach towards the construction of effective field theories with linearly realised \Poincare symmetries and additional non-linear symmetries. Our results can be phrased in the following way. 

We begin with space-time itself. The starting point is space-time translations with generators $P_\mu$ to which we associate the space-time coordinates $x^\mu$. These can be seen as the essential Goldstones for the coset construction of space-time itself, transforming non-linearly under translations with a constant shift. An additional symmetry structure can be added, namely, Lorentz generators whose Goldstones can be seen as inessential since they are related to translations via $[P,M] = P$. This additional symmetry does therefore not introduce new associated coordinates but is instead realised as a higher-order transformation on the original ones. Moreover, since $[M,M] = M$ this is an exceptional algebra in the sense defined in section \ref{Spacetime} and it does not allow for additional non-linear generators, thus completing the \Poincare space-time structure.

In order to build a field theory, one has to include additional generators associated with new coset coordinates $\phi(x)$ (with any Lorentz structure suppressed). As outlined in section \ref{Spacetime}, any additional symmetries realised on the same field necessarily build up an inverse Higgs tree via commutators with space-time translations, of the form $[ P_\mu, G_i ] = G_{i-1}$ with the possible non-linear generators fixed by the Taylor expansion of the essential Goldstone. The requirement of a canonical propagator restricts the possible symmetries at every order of the inverse Higgs tree and for scalars and fermions reduces to a single Lorentz representation, and thus a unique structure at every order.

The final step towards finding exceptional algebras with highly constrained EFTs, then consists of a (model-dependent) classification of possible commutators between different non-linear generators. We have performed this classification in full generality for multiple scalars and for multiple fermions. Assuming a shift symmetry at lowest order\footnote{If instead one opts for a dilatation symmetry at lowest order this results uniquely in conformal algebras.}, the only exceptional possibilities for a single scalar turn out to be the five-dimensional \Poincare algebra and the Special Galileon algebra. Turning to multiple scalars, one encounters higher-dimensional \Poincare algebras as well as trivial couplings between DBI scalars (or a Special Galileon scalar) and shift-symmetric axions (or indeed any scalars which realise an internal symmetry group). We have also seen that one can couple DBI scalars to Galileons. For fermions, the only non-trivial algebra that can be constructed is the super-\Poincare one (coupled to an arbitrary number of shift-symmetric fermions). We have proven these statements to arbitrary high order in the inverse Higgs trees. Remarkably, in all cases the commutators between non-linear generators produce \Poincare generators, instead of non-linear generators themselves, exactly mirroring the symmetric structure of space-time discussed above.

In the final subsection we have also addressed the possible couplings between two different \Poincare representations: a scalar and a vector. As would be expected, this includes the coupling between the Goldstone scalar of an exceptional algebra and a gauge vector transforming as a matter field. For the Special Galileon case, we anticipate this to be relevant for the scalar-vector interactions and soft limits identified in \cite{SoftBootstrap}.

These results underline the generality of the algebraic approach, with the structure of the inverse Higgs tree plus the invariance of the kinetic terms giving rise to an unambiguous series of inessential generators. This makes the full problem amenable to a general analysis, as illustrated by the specific examples that we have fully analysed. Similar analyses can be performed for different \Poincare representations, or couplings between different irreps. 

Moreover, the outlined approach is by no means specific to the case of \Poincare invariant field theories, and can {\it mutatis mutandis} be applied to theories with a different linear symmetry. 
An example of this would be SUSY EFTs, with a linearly realised super-Poincare symmetry.  
In $\mathcal{N} = 1$ superspace, each non-linear generator creates a Goldstone superfield. Once again, there can be degeneracy between the various Goldstone excitations. On top of the ordinary inverse Higgs constraints, which relate some Goldstone modes to the space-time derivatives of others, there are new \emph{fermionic} inverse Higgs constraints. These relate some Goldstone superfields to the superspace derivatives of others. The basic algebraic structure that underlies fermionic inverse Higgs constraints is $[Q_\alpha, G_i] = G_{i - 1} + \ldots$, where $Q_{\alpha}$ are the fermionic supertranslations. Once again, this simple algebraic structure makes it possible to perform an exhaustive classification of exceptional EFTs. We will present our results for linear supersymmetry in a forthcoming companion paper \cite{SUSY-EFTs}.

Other contexts where this analysis could prove useful are time-dependent backgrounds as considered in cosmology, initiated in \cite{SymmetricSuperfluids}, as well as non-relativistic settings as considered in condensed matter physics, where it is also natural to wonder about the role of Goldstone modes, their soft-limits and the most general non-linear symmetries.

\subsection*{Acknowledgements}

It is a pleasure to thank Brando Bellazzini, Eric Bergshoeff, James Bonifacio, Kurt Hinterbichler, Callum Jones, Enrico Pajer and Yusuke Yamada for very useful discussions. We acknowledge the Dutch funding agency ``Netherlands Organisation for Scientific Research'' (NWO) for financial support.

\end{document}